\documentclass[twocolumn, twocolappendix]{aastex63}
\pdfoutput=1

\received{2021 December 15}
\revised{2022 February 14}
\accepted{2022 March 7}
\submitjournal{ApJ}

\shorttitle{Ly$\alpha$ Emission in Low-$z$ WLQs}
\shortauthors{Paul et al.}

\graphicspath{{./}}
\usepackage{multirow}
\usepackage{amsmath}

\defcitealias{DS09}{DS09}
\defcitealias{Plotkin15}{P15}

\begin{document}

\title{Connecting Low- and High-Redshift Weak Emission-Line Quasars via HST Spectroscopy of Ly$\alpha$ Emission}

\correspondingauthor{Jeremiah D. Paul} 
\email{jdpaul@nevada.unr.edu}

\author[0000-0003-0040-3910]{Jeremiah D. Paul}
\affiliation{Department of Physics, University of Nevada, Reno, NV 89557, USA}

\author[0000-0002-7092-0326]{Richard M. Plotkin}
\affiliation{Department of Physics, University of Nevada, Reno, NV 89557, USA}

\author[0000-0003-4327-1460]{Ohad Shemmer}
\affiliation{Department of Physics, University of North Texas, Denton, TX 76203, USA}

\author{Scott F. Anderson}
\affiliation{Department of Astronomy, University of Washington, Box 351580, Seattle, WA 98195, USA}

\author[0000-0002-0167-2453]{W. N. Brandt}
\affiliation{Department of Astronomy \& Astrophysics, 525 Davey Lab, The Pennsylvania State University, University Park, PA 16802, USA}
\affiliation{Institute for Gravitation and the Cosmos, The Pennsylvania State University, University Park, PA 16802, USA}
\affiliation{Department of Physics, 104 Davey Lab, The Pennsylvania State University, University Park, PA 16802, USA}

\author[0000-0003-3310-0131]{Xiaohui Fan}
\affiliation{Steward Observatory, University of Arizona, 933 N. Cherry Ave, Tucson, AZ 85721, USA}

\author[0000-0001-5802-6041]{Elena Gallo}
\affiliation{Department of Astronomy, University of Michigan, 1085 S. University Ave, Ann Arbor, MI 48109, USA}

\author[0000-0002-9036-0063]{Bin Luo}
\affiliation{School of Astronomy and Space Science, Nanjing University, Nanjing, Jiangsu 210093, China}
\affiliation{Key Laboratory of Modern Astronomy and Astrophysics (Nanjing University), Ministry of Education, Nanjing, Jiangsu 210093, China}

\author[0000-0002-8577-2717]{Qingling Ni}
\affiliation{Institute for Astronomy, University of Edinburgh, Royal Observatory, Edinburgh EH9 3HJ, UK}
\affiliation{Department of Astronomy \& Astrophysics, 525 Davey Lab, The Pennsylvania State University, University Park, PA 16802, USA}
\affiliation{Institute for Gravitation and the Cosmos, The Pennsylvania State University, University Park, PA 16802, USA}

\author[0000-0002-1061-1804]{Gordon T. Richards}
\affiliation{Department of Physics, Drexel University, 32 S. 32nd Street, Philadelphia, PA 19104, USA}

\author[0000-0001-7240-7449]{Donald P. Schneider}
\affiliation{Department of Astronomy \& Astrophysics, 525 Davey Lab, The Pennsylvania State University, University Park, PA 16802, USA}
\affiliation{Institute for Gravitation and the Cosmos, The Pennsylvania State University, University Park, PA 16802, USA}

\author[0000-0001-7349-4695]{Jianfeng Wu}
\affiliation{Department of Astronomy, Xiamen University, Xiamen, Fujian 361005, China}

\author[0000-0001-9314-0552]{Weimin Yi}
\affiliation{Yunnan Observatories, Kunming 650216, China}
\affiliation{Department of Astronomy \& Astrophysics, 525 Davey Lab, The Pennsylvania State University, University Park, PA 16802, USA}
\affiliation{Key Laboratory for the Structure and Evolution of Celestial Objects, Chinese Academy of Sciences, Kunming 650216, China}

\begin{abstract}
We present ultraviolet spectroscopy covering the Ly$\alpha$ + \ion{N}{5} complex of six candidate low-redshift ($0.9 < z < 1.5$) weak emission-line quasars (WLQs) based on observations with the Hubble Space Telescope. The original systematic searches for these puzzling Type 1 quasars with intrinsically weak broad emission lines revealed an $N \approx 100$ WLQ population from optical spectroscopy of high-redshift ($z>3$) quasars, defined by a Ly$\alpha$ + \ion{N}{5} rest-frame equivalent width (EW) threshold $< 15.4$~{\AA}. Identification of lower-redshift ($z<3$) WLQ candidates, however, has relied primarily on optical spectroscopy of weak broad emission lines at longer rest-frame wavelengths. With these new observations expanding existing optical coverage into the ultraviolet, we explore unifying the low- and high-$z$ WLQ populations via EW[Ly$\alpha$+\ion{N}{5}]. Two objects in the sample unify with high-$z$ WLQs, three others appear consistent with the intermediate portion of the population connecting WLQs and normal quasars, and the final object is consistent with typical quasars. The expanded wavelength coverage improves the number of available line diagnostics for our individual targets, allowing a better understanding of the shapes of their ionizing continua. The ratio of EW[Ly$\alpha$+\ion{N}{5}] to EW[\ion{Mg}{2}] in our sample is generally small but varied, favoring a soft ionizing continuum scenario for WLQs, and we find a lack of correlation between EW[Ly$\alpha$+\ion{N}{5}] and the X-ray properties of our targets, consistent with a ``slim-disk" shielding gas model. We also find indications that weak absorption may be a more significant contaminant in low-$z$ WLQ populations than previously thought.
\end{abstract}


\section{Introduction} \label{sec:intro}

Supermassive black holes (SMBHs; $ 10^7 \lesssim M_{\rm BH} \lesssim 10^{9}~M_\odot$) reside in the nuclei of all large galaxies, and it is likely that every SMBH goes through at least one luminous quasar phase where it accretes at $\gtrsim$10\% of its Eddington luminosity ($L_{\rm Edd}$; e.g., \citealt{Soltan82,Richstone98,Kormendy13}). In the standard unification paradigm \citep[e.g.,][]{Antonucci93,Urry95}, one of the most prominent components of unobscured (i.e., Type 1) quasars is the broad emission line region (BELR), where gas embedded deep within the gravitational potential well of the SMBH reprocesses photons from the accretion disk and X-ray corona into Doppler-broadened line emission \citep[e.g.,][]{Rees84,Osterbrock86,Peterson93,Bentz09}. These broad lines are observed with typical FWHM $\gtrsim10^3$ km s$^{-1}$ in rest-frame optical and ultraviolet (UV) spectroscopy \citep[e.g.,][]{Osterbrock82}. 

Examples of Type 1 quasars with unusually weak or missing broad emission lines have emerged over the last $\sim$25 years \citep[e.g.,][]{McDowell95, Fan99, Anderson01, Leighly07a}. Using the Sloan Digital Sky Survey \citep[SDSS;][]{SDSS1}, \citet[][hereafter \citetalias{DS09}]{DS09} performed the first systematic search for these weak-lined quasars (WLQs), revealing the existence of a population of such objects. \citetalias{DS09} measured the rest-frame equivalent width (EW) of the Ly$\alpha$ $\lambda$1216 + \ion{N}{5} $\lambda$1240 complex in $\sim$5000 high-redshift ($z>3$) quasars. Finding that the general population follows a log-normal distribution in EW[Ly$\alpha$+\ion{N}{5}] except for a $>3\sigma$-weak tail with an excess of $\sim$100 objects, they defined WLQs as having EW[Ly$\alpha$+\ion{N}{5}] $<15.4$~{\AA}. They also identified a similar excess in the $>3\sigma$-weak tail of the distribution of the \ion{C}{4} $\lambda$1549 broad emission line, corresponding to EW[\ion{C}{4}] $<10$~{\AA}. The weak BELR in WLQs is very likely an intrinsic feature, i.e., it is not solely an artifact of orientation-induced obscuration or absorption (e.g., \citealt{Anderson01}; \citetalias{DS09}), gravitational lensing or microlensing (e.g., \citealt{Shemmer06}; \citetalias{DS09}), and/or Doppler boosting (e.g., \citealt{Plotkin10a,Lane11}). 

WLQs are usually observed to have optical$-$UV continuum shapes similar to those of typical quasars. However, their broad emission lines of high ionization potential (e.g., \ion{C}{4}) appear to be preferentially weakened compared to lines of low ionization potential (e.g., H$\beta$), which suggests their BELR gas is placed in an unusual ionization state by exposure to an abnormally soft photoionizing continuum (e.g., \citealt{Dietrich02}; \citealt{Plotkin15}, hereafter \citetalias{Plotkin15}). At the same time, WLQs are also known to exhibit unusual X-ray properties. A relatively large fraction of WLQs ($\sim$50\%) are X-ray weak compared to typical quasars ($\lesssim$6\%), and they can extend to more extreme values of X-ray weakness \citep[e.g.,][]{Ni18,Pu20}. X-ray normal WLQs generally show steep power-law X-ray spectra with a soft excess, suggesting high accretion rates \citep[e.g.,][]{Luo15,Marlar18}. On the other hand, the population of X-ray weak WLQs appears to have hard X-ray spectra on average, indicating likely absorption and/or reflection \citep[e.g.,][]{Wu12,Luo15}. Furthermore, while the normal radio-quiet quasar population displays a correlation between X-ray weakness and lower EW[\ion{C}{4}] \citep{Gibson08}, this correlation does not appear to extend to the WLQ population \citep{Ni18,Ni22,Timlin20}.

Several physical interpretations have been proposed to explain the above observational characteristics of WLQs. Presently, the foremost is that of a column of ``shielding" gas existing between the X-ray corona and BELR \citep[e.g.,][]{Wu11,Wu12,Luo15}, likely related to the inner edge of an optically- and geometrically-thick, super-Eddington ``slim" accretion disk \citep[e.g.,][]{Abramowicz88,Czerny19} and its wide-angle outflows \citep[e.g.,][]{Murray95,Castello-Mor17,Jiang19,Giustini19}. This configuration is expected to shield the BELR from X-ray and extreme UV radiation emitted by the corona and innermost regions of the accretion disk, softening the incident ionizing continuum. The varied X-ray properties we observe are then explained as a byproduct of orientation, as the edge of the disk and the outflows will obscure X-ray emission from our line of sight at larger inclination angles (\citealt{Luo15,Ni22}; see also Fig. 1 of \citealt{Ni18}).

Alternative explanations have been suggested for the WLQ phenomenon that we describe below for completeness, but we stress that none can also explain the unusual X-ray properties of WLQs (unless WLQs represent a heterogeneous population of objects, and/or multiple mechanisms contribute to their broad line weakness). Other ways to produce preferentially weaker high-ionization potential emission lines include WLQs being powered by exceptionally massive accreting SMBHs ($M_{\rm BH} > 3.6 \times 10^9 M_{\odot}$ for non-spinning black holes), which could produce cooler accretion disks with softer ionizing spectra \citep{Laor11}, or super-Eddington accretion producing weaker X-ray coronae, thereby yielding softer, UV-peaked spectral energy distributions (SEDs) \citep[e.g.,][]{Leighly07a,Leighly07b}. In other cases, it has been suggested that the BELR is gas deficient \citep[e.g.,][]{Shemmer10,Hryniewicz10}.

The large sample of WLQs from \citetalias{DS09} consists exclusively of high redshift ($z>3$) objects. This bias is simply an artifact of the SDSS optical spectral range. Nevertheless, WLQ candidates have also been identified at lower redshifts ($z<3$), but with the caveat that their selection is based on weak broad emission lines at longer wavelengths, such as \ion{C}{4}, \ion{C}{3}] $\lambda$1909, and/or \ion{Mg}{2} $\lambda$2800 \citep[e.g.,][]{Collinge05,Hryniewicz10,Plotkin10a,Plotkin10b,Meusinger14}. This redshift-based division of the WLQ population poses two problems. First, the lack of a universal standard for classification hampers our ability to confidently unify the two populations \citep[e.g.,][]{Nikolajuk12,Wu12,Luo15}. Second, given that the BELR is expected to produce broad emission lines across the entire rest-frame optical$-$UV range, failure to capture that full range in individual objects prevents the use of diagnostics critical for discriminating between different models for broad line weakness.

Here, we present a pilot study using the Space Telescope Imaging Spectrograph (STIS; \citealt{HST_STIS}) aboard the Hubble Space Telescope (HST) to extend existing SDSS optical coverage of six candidate low-redshift WLQs ($1 \lesssim z \lesssim 1.5$) into the UV. Our primary objective is to compare EW[Ly$\alpha$+\ion{N}{5}] between low- and high-redshift WLQs and assess whether the two populations can indeed be unified. Expanding the wavelength coverage for these six individual objects also increases the number of available line diagnostics, allowing a better understanding of the shapes of their ionizing continua. 

The paper is organized as follows: In Section \ref{sec:obs}, we describe our HST STIS observations and data reduction. Section \ref{sec:spec} covers our spectral analysis, including measurement of the UV continuum shape and emission line strengths. We present our results in Section \ref{sec:results} and discuss them in the contexts of the parent quasar and high-redshift WLQ populations in Section \ref{sec:discussion}. Finally, we summarize our conclusions in Section \ref{sec:summ}. Throughout, we use the term ``quasar" to specifically denote Type 1 quasars. We adopt the following cosmology: $H_0=67.4$ km s$^{-1}$ Mpc$^{-1}$, $\Omega_{\rm M}=0.315$, and $\Omega_{\rm \Lambda}=0.685$ \citep{Planck20}.

\section{Hubble Space Telescope Observations} \label{sec:obs}

\subsection{Sample Selection} \label{subsec:samp}

\begin{figure*}[tb!]
\gridline{\fig{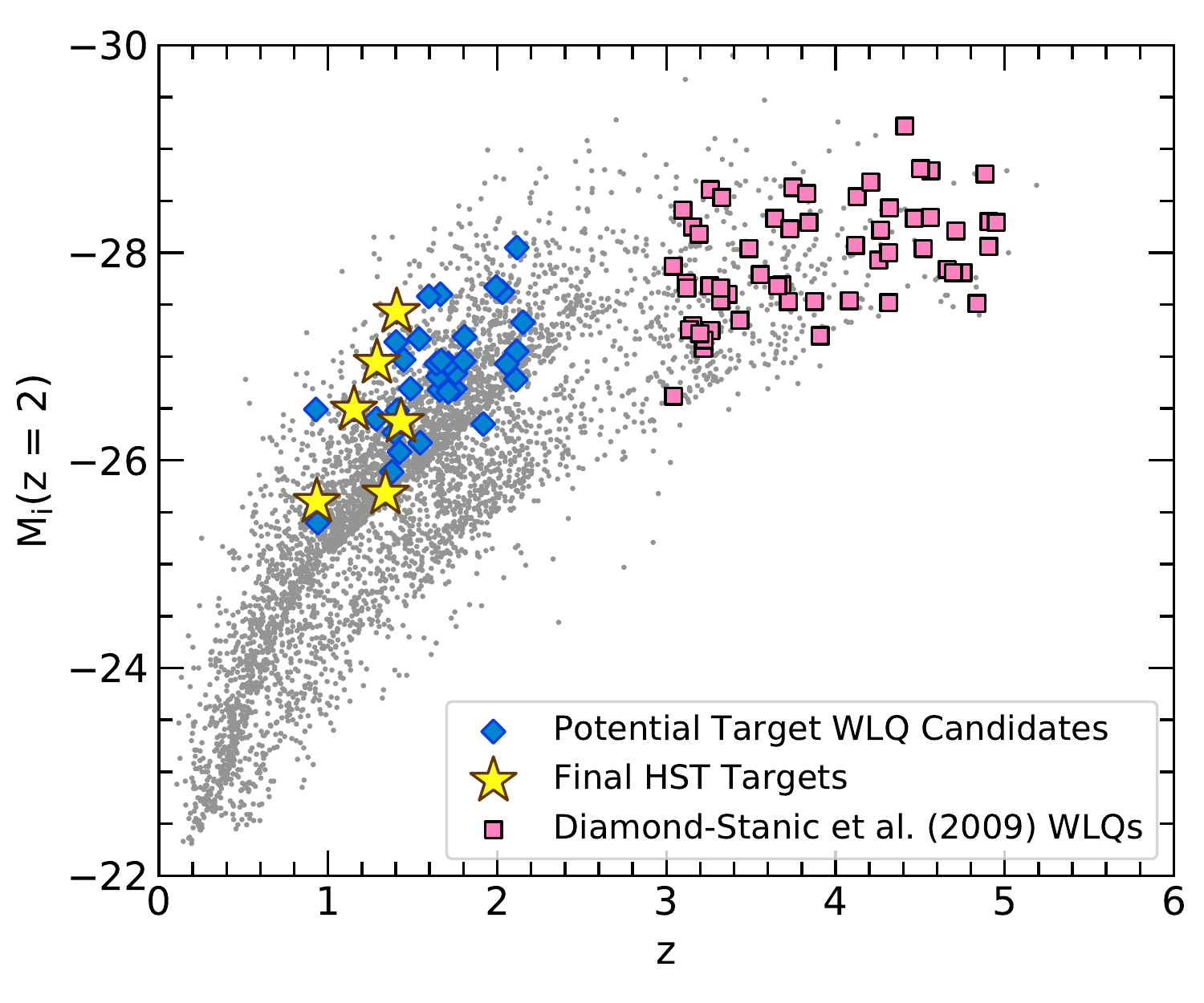}{0.53\textwidth}{(a)}
          \fig{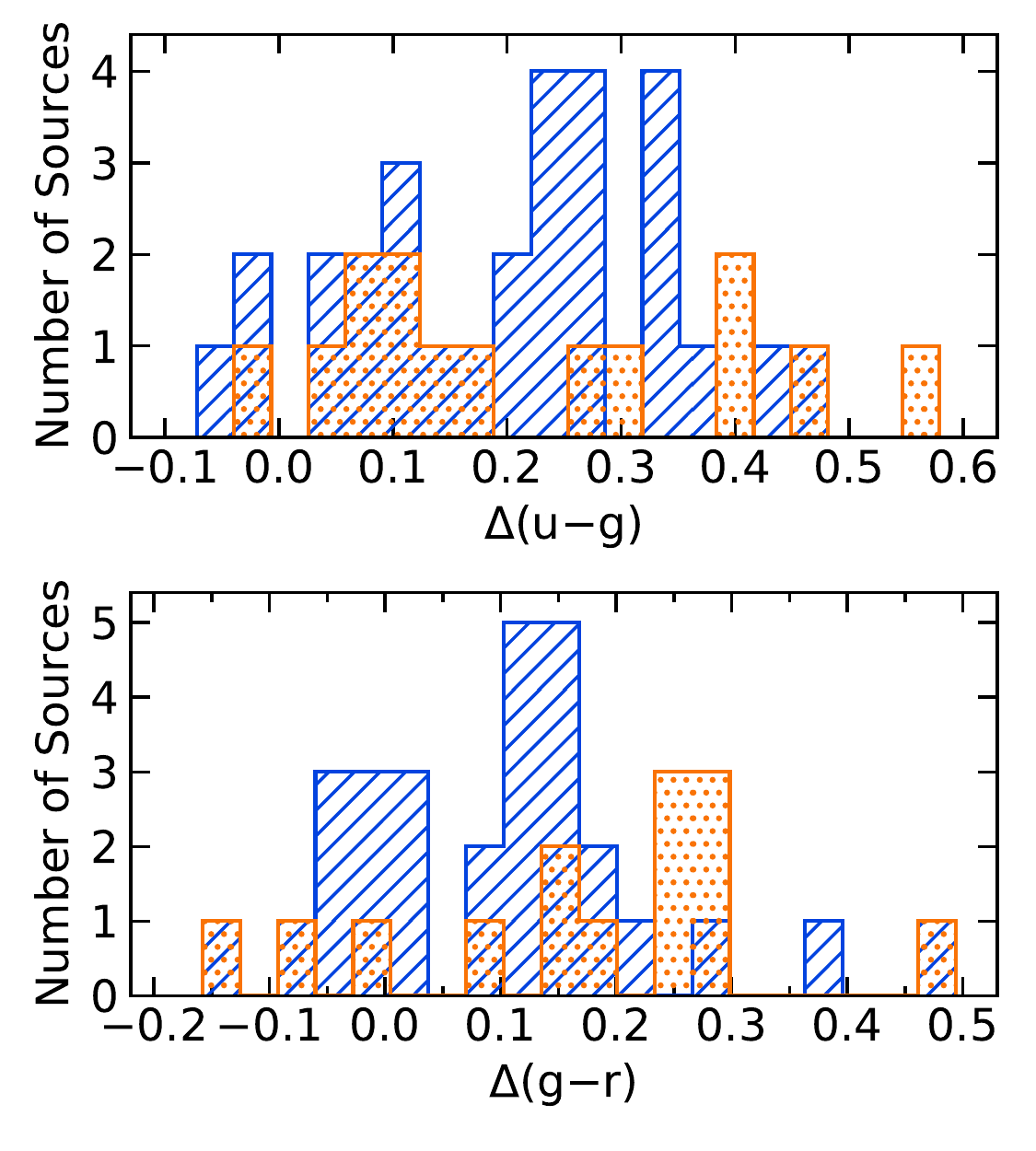}{0.39\textwidth}{(b)}}
\caption{(a) SDSS \textit{i}-band absolute magnitude, $M_i$ (at $z=2$; see \citealt{Richards06}) vs.\ redshift, $z$. Blue diamonds represent our initial sample of 43 potential target low-$z$ WLQ candidates, and yellow stars represent the final six targets selected for HST observation (see Section \ref{subsec:samp}). For comparison, pink squares show the high-$z$ WLQ catalog of \citetalias{DS09}. Grey dots represent the 105,783 objects (via a random sample of 5,000) in the full SDSS DR7 quasar catalog \citep[][]{Schneider10}. (b) Histograms comparing the relative SDSS $\Delta(u-g)$ (top panel) and $\Delta(g-r)$ (bottom panel) color distributions of GALEX-selected (blue hatched bins) vs.\ rejected (orange dotted bins) objects from our initial sample of 43 potential targets. We find no statistical difference between the GALEX-selected and rejected distributions (see Section \ref{subsec:samp}). \label{fig:color_hist}}
\end{figure*}

We identified potential HST targets by examining lists of $z<3$, weak-featured quasars that were serendipitously recovered during searches for BL Lac objects in SDSS data releases \citep[$\sim$$10^2$ objects;][]{Collinge05,Anderson07,Plotkin10b}. These quasars passed optical spectroscopic criteria to be classified as BL Lac objects, i.e., all observed emission lines had EW $<5$~{\AA},\footnote{Note that \citet{Plotkin10b} did not account for blended \ion{Fe}{2} and \ion{Fe}{3} emission when performing their EW measurements, such that some unbeamed quasars in their sample will have broad lines with EW $>$5~{\AA} (especially \ion{Mg}{2}) when properly accounting for blended iron emission.} and the 4000~{\AA} break, if present, was smaller than $40\%$ (see \citealt{Plotkin10b} for details). However, all objects also had faint radio detections or upper limits that firmly placed them in the radio-quiet regime,\footnote{We adopt the standard definition whereby radio-quiet quasars have radio-to-optical flux density ratios $R=f\textsubscript{5 GHz}/f\textsubscript{4400 {\AA}} < 10$, where $f\textsubscript{\rm 5 GHz}$ and $f\textsubscript{4400 {\AA}}$ are, respectively, the radio and optical flux densities at 5~GHz and 4400~{\AA} (\citealt{Kellermann89,Stocke92}).} based on data from the Faint Images of the Radio Sky at Twenty cm survey (FIRST; \citealt{Becker95}) and the NRAO VLA Sky Survey (NVSS; \citealt{Condon98}). To exclude targets that might have weakly-beamed continua, we only considered objects with radio loudnesses $R<8$, which represents a $>3\sigma$ departure from the expected radio loudness of normal SDSS BL Lac objects (e.g., \citealt{Plotkin10b}). We then restricted ourselves to objects with secure redshifts from weak emission features, which left 43 potential targets. The redshifts and SDSS $i$-band magnitudes of these 43 low-$z$ WLQ candidates are shown relative to both the parent SDSS quasar population and high-$z$ WLQs from \citetalias{DS09} in Figure~\ref{fig:color_hist}a.

To select targets for UV observations with HST, we correlated these 43 objects to the GALEX UV catalogs \citep{Bianchi11,Bianchi17}. Using a 3 arcsecond match radius and visually examining the GALEX images for each match, we identified 29 sources with suitable GALEX counterparts. The remaining 14 objects were rejected, comprising six matched objects with a GALEX ``artifact" or ``extractor" flag, two objects lacking a GALEX detection within the match radius, and six objects with locations not covered by GALEX. To determine whether these cuts bias the sample toward objects with bluer continua in their SDSS spectra, we performed Kolmogorov–Smirnov (K-S) and Anderson-Darling (A-D) tests comparing the distributions of the 29 ``GALEX-selected" vs.\ 14 ``rejected" objects in relative SDSS optical $\Delta(u-g)$ and $\Delta(g-r)$ colors.\footnote{The mean redshift of the ``rejected" population ($\langle z \rangle = 1.71$) is slightly higher than that of the ``selected" population ($\langle z \rangle = 1.44$). We accounted for this by comparing relative colors, e.g., $\Delta(u-g)$, defined as the difference between the observed color, $u-g$, and the median color at redshift, $\langle u-g \rangle$, from \citet{Richards01}. Note also that the makeup of the rejected sample at this point in the selection process precludes testing GALEX-SDSS colors (e.g., ${\rm NUV}-g$).} The results of these tests are in good agreement with the null hypothesis: for $\Delta(u-g)$, $p=0.69$ and 0.25 from the K-S and A-D tests, respectively, and for $\Delta(g-r)$, $p=0.04$ (K-S) and 0.05 (A-D); see also Figure \ref{fig:color_hist}b.

We next restricted the 29-object GALEX-selected sample to include only objects with $0.9 < z < 1.5$ and GALEX NUV apparent magnitudes $m_{\rm NUV} < 19.1$. These cuts were necessary for economical HST observations, allowing us to obtain sufficient signal-to-noise ($>10$ per resolution element) near the continuum of the Ly$\alpha$ + \ion{N}{5} emission complex in only 1$-$2 HST orbits per target.\footnote{This final criterion introduced an unavoidable bias toward objects with bluer GALEX-SDSS colors; further tests of our targets' optical$-$UV continua are explored in Section \ref{subsec:results_absmag}.} After applying the above cuts, we were left with a sample of six candidate WLQs (see Table \ref{tab:obslog}). All six objects have roughly similar optical$-$UV luminosities ($l$\textsubscript{2500 {\AA}} $\sim 10^{31}$ erg s$^{-1}$ Hz$^{-1}$). Furthermore, while we did not utilize X-ray emission as an explicit selection criterion, all of our candidates have been observed by the Chandra X-ray Observatory \citep{Wu12,Luo15}, and they fortuitously span nearly the full range of X-ray to optical flux ratios so far observed for WLQs. Throughout this work, we refer to each object by its SDSS designation truncated to the first four digits. All of our targets but one (SDSS J1447) lack coverage of [\ion{O}{3}] or H$\beta$ emission, so for consistency we simply adopt the values from \citet{Hewett10} as ``systemic" redshifts. We do not attempt to measure redshift from Ly$\alpha$ emission as it is expected to be weak/absent in our targets.

\subsection{Observations and Data Reduction} \label{subsec:reduc}

Observations with STIS were completed under HST program GO-13298 (PI Plotkin) in Cycle 21, using the NUV-MAMA detector with the G230L grating ($R\approx 500$, dispersion $\Delta\lambda = 1.58$~{\AA} pixel$^{-1}$) and $52''\times0.2''$ slit. One target (SDSS J1447) required two orbits, while the other five targets were each observed in a single orbit. Two exposures per orbit were taken, dithered along the slit in order to correct for cosmic rays and hot pixels. Table \ref{tab:obslog} summarizes the targets and observations.

\begin{deluxetable*}{lcccccrcc}[tb!]
\tablenum{1}
\tablecaption{Sample Properties and HST STIS Observation Log\label{tab:obslog}}
\tablehead{
\colhead{Source Name} & \colhead{$z$} & \colhead{$m_{\rm NUV}$} & \colhead{$\log l$\textsubscript{2500 {\AA}}} & \colhead{$L_{\rm bol}/L_{\rm Edd}$} & \colhead{$\alpha_{\rm ox}$} & \colhead{$\Delta\alpha_{\rm ox}$} & \colhead{Obs. Date} & \colhead{Exp.\ Time}\\
\colhead{(SDSS J)} & \colhead{} & \colhead{(mag)} & \colhead{(erg s$^{-1}$ Hz$^{-1}$)} & \colhead{} & \colhead{} & \colhead{} & \colhead{} & \colhead{(s)}\\
\colhead{(1)} & \colhead{(2)} & \colhead{(3)} & \colhead{(4)} & \colhead{(5)} & \colhead{(6)} & \colhead{(7)} & \colhead{(8)} & \colhead{(9)}
}
\startdata
081250.80+522530.8 & 1.153 & 19.13 & 30.90 & $0.23$ & $-$2.03\textsuperscript{a} & $-$0.41 & 2014 Sep 04 & 2446\\
090843.25+285229.8 & 0.933 & 19.09 & 30.55 & $0.17$ & $-$1.46\textsuperscript{b} & 0.11 & 2014 Feb 26 & 2250\\
125219.48+264053.9 & 1.288 & 18.76 & 31.06 & $0.11$ & $-$2.03\textsuperscript{a} & $-$0.39 & 2014 Apr 24 & 2270\\
144741.76$-$020339.1 & 1.431 & 19.11 & 30.97 & $1.33$ & $-$1.76\textsuperscript{b} & $-$0.13 & 2014 Aug 22 & 3593\\
153044.08+231013.5 & 1.406 & 18.76 & 31.25 & $0.32$ & $-$1.45\textsuperscript{a} & 0.22 & 2014 Mar 03 & 2260\\
162933.60+253200.6 & 1.340 & 18.76 & 30.71 & $0.03$ & $-$1.62\textsuperscript{b} & $-$0.03 & 2014 May 08 & 2230\\
\enddata
\tablecomments{Column (1): object name. Column (2): redshift from \citet{Hewett10}. Column (3): GALEX NUV apparent magnitude \citep[][]{Bianchi11,Bianchi17}. Column (4): base-10 logarithm of the 2500~{\AA} specific luminosity, $l$\textsubscript{2500 {\AA}}$= 4 \pi D^{2}_{\rm L} f$\textsubscript{2500 {\AA}}$ / (1 + z)$, where $f$\textsubscript{2500 {\AA}} is the flux density at 2500~{\AA} from \citet{Shen11} and $D_{\rm L}$ is the luminosity distance found using the \texttt{astropy.cosmology} package \citep[][]{astropy13,astropy18}. Column (5): Eddington ratio (\ion{Mg}{2}-based estimate) from \citet{Shen11}, except for SDSS J1447 (H$\beta$-based estimate) from \citetalias{Plotkin15}. Column (6): optical$-$UV to X-ray spectral slope $\alpha_{\rm ox} = $ 0.3838 log$(f_{\rm 2 keV} / f$\textsubscript{2500 {\AA}}) (see Section \ref{subsec:results_x-ray}). Column (7): X-ray weakness parameter $\Delta \alpha_{\rm ox} = \alpha_{\rm ox} - \alpha_{\rm ox,qso}$, where $\alpha_{\rm ox,qso}$ is the value predicted by the $\alpha_{\rm ox}$$-$$l\textsubscript{2500 {\AA}}$ anticorrelation displayed by broad-line radio-quiet quasars (e.g., \citealt{Steffen06,Just07,Timlin20}). For consistency with prior WLQ studies, we adopt the best-fit relationship given by \citet{Just07}. Column (8): HST observation date. Column (9): total HST observation exposure time.\\
\textsuperscript{a}\,Found using $f_{\rm 2 keV}$ from \citet{Wu12}.\\
\textsuperscript{b}\,Found using $f_{\rm 2 keV}$ from \citet{Luo15}.
}
\end{deluxetable*}

\begin{figure*}[tb!]
\plotone{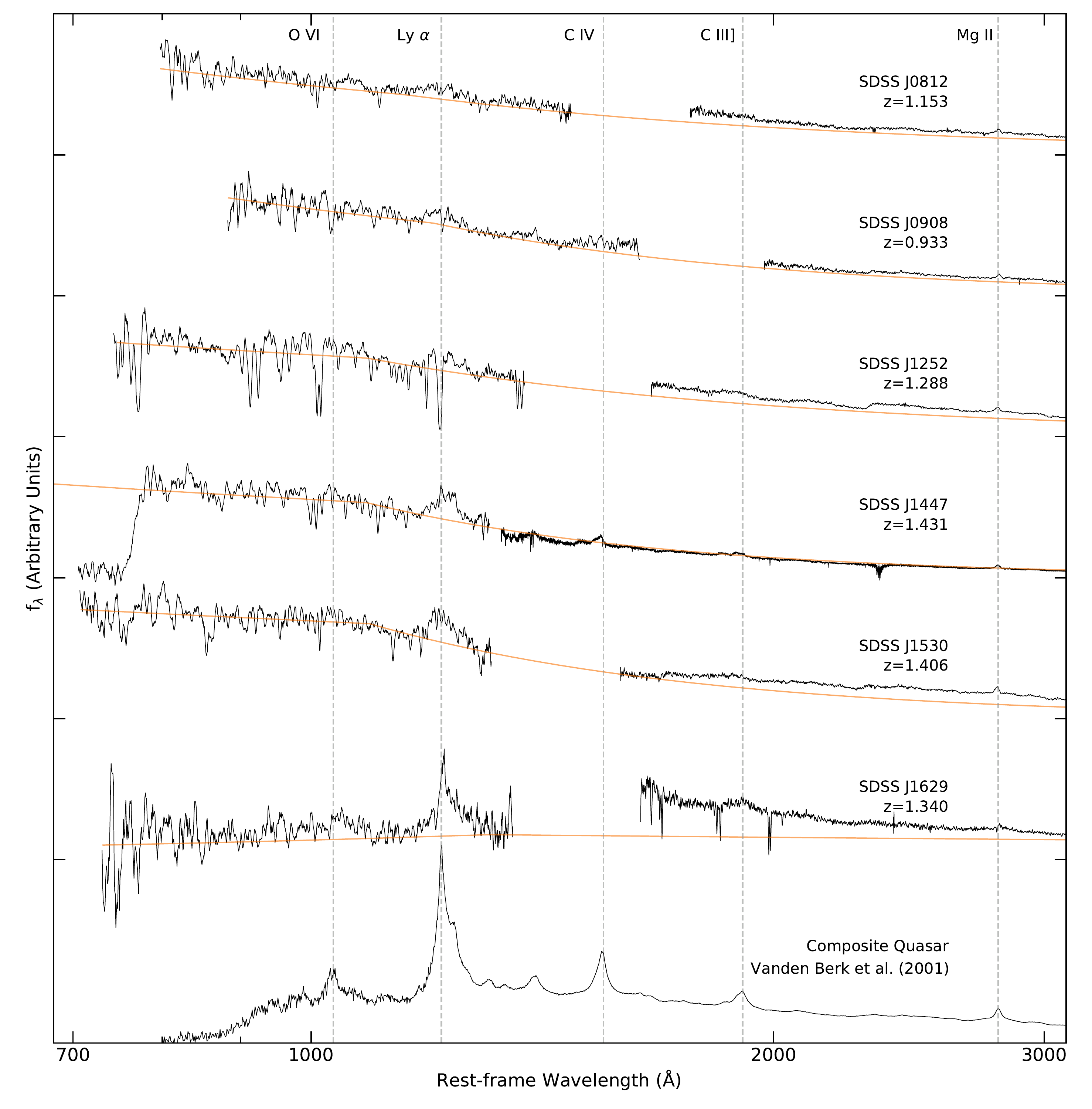}
\caption{UV spectra redward to 3100~{\AA} for all six target WLQ candidates. The gaps are breaks in spectral coverage; left of the gap is the HST STIS spectrum for all, and right of the gap is the SDSS spectrum for all but SDSS J1447. For SDSS J1447, we instead show the X-shooter spectrum from \citetalias{Plotkin15} in order to illustrate its \ion{C}{4} emission. The bottom-most spectrum is the quasar composite from \citet{VB01}, shown for comparison. Observed flux density is plotted in solid black, and the fitted power-law continuum (See Section \ref{subsec:cont}) is plotted in orange. The vertical axis is the flux density ($f_{\lambda}$) in arbitrary linear units, with ticks denoting the zero-flux level for each spectrum. Our targets have been individually scaled on this axis to display as much detail as possible, while the \citet{VB01} spectrum is scaled to have its $f_{\lambda}$ level at 1450~{\AA} roughly match that of SDSS J0908. The horizontal axis gives the rest-frame wavelength in {\AA} on a log$_{10}$ scale. Rest-frame wavelength positions of several prominent quasar emission lines are indicated by vertical dashed lines with labels at figure top. The spectra have been smoothed via 1D box convolution (5 pixel width). Note: SDSS J1447 shows a prominent \ion{Mg}{2} absorption feature at $\lambda_{\rm rest} \approx 2340$~{\AA}. The apparent strength of this feature is likely enhanced due to noise, as in the observed frame it falls near the UVB/VIS break in X-shooter coverage. \label{fig:all_spectra}}
\end{figure*}

Calibrated spectra (at each dither position) were downloaded from the Mikulski Archive for Space Telescopes, which used the {\tt calstis} v3.4 reduction pipeline. Further processing was performed using the {\tt STSDAS} package in {\tt PyRAF}. We first removed the dither offsets using the task {\tt sshift}, and then combined each sub-exposure using {\tt mscomb} to create a single spectrum per object. Finally, one-dimensional spectra were extracted using the task {\tt x1d} with an 11-pixel aperture, which were then corrected for Galactic extinction using the \citet{Schlafly11} recalibration of the \citet{Schlegel98} maps and adopting a \citet{Cardelli89} reddening law. The resulting HST UV spectra are presented in Figure \ref{fig:all_spectra}.

\section{Analysis} \label{sec:spec}

\begin{deluxetable*}{lccccc}[htb!]
\tablenum{2}
\tablecaption{Fitting Parameters and Measurements of Broken Power Law Continua\label{tab:windows}}
\tablehead{
\colhead{Source Name} & \colhead{Fit Windows\textsuperscript{a}} & \colhead{$\lambda_{\rm break}$} & \colhead{$\alpha_{\rm FUV}$} & \colhead{$\alpha_{\rm NUV}$} & \colhead{Comments}\\
\colhead{(SDSS)} & \colhead{} & \colhead{({\AA})} & \colhead{} & \colhead{} & \colhead{}\\
\colhead{(1)} & \colhead{(2)} & \colhead{(3)} & \colhead{(4)} & \colhead{(5)} & \colhead{(6)}
}
\startdata
\multirow{2}*{J0812} & \multirow{2}*{A, D, H, L, O, R, T} & \multirow{2}*{$1145.7^{+3.0}_{-195.7}$} & \multirow{2}*{$-1.0 \pm 0.2$} & \multirow{2}*{$-1.4^{+0.3}_{-0.1}$} & \multirow{2}{0.33\textwidth}{NAL system at $z\approx0.79$: \ion{Mg}{2} $\lambda2800$, \ion{C}{4} $\lambda1549$, Ly$\alpha$, \ion{O}{6} $\lambda1035$, Ly$\beta$.\textsuperscript{b}}\\
 & & & & & \\
\hline
\multirow{2}*{J0908} & \multirow{2}*{D, H, L, O, R, T} & \multirow{2}*{$1200.0^{+62.4}_{-220.1}$} & \multirow{2}*{$-1.0^{+0.4}_{-0.3}$} & \multirow{2}*{$-2.0^{+0.3}_{-0.2}$} & \multirow{2}{0.33\textwidth}{\nodata}\\
 & & & & & \\
\hline
\multirow{2}*{J1252} & \multirow{2}*{A, D, I, M, O, R, S} & \multirow{2}*{$1086.2^{+6.3}_{-35.9}$} & \multirow{2}*{$-0.5 \pm 0.2$} & \multirow{2}*{$-1.5 \pm 0.2$} & \multirow{2}{0.33\textwidth}{NAL system at $z\approx1.24$: \ion{Si}{4} $\lambda1398$, Ly$\alpha$, Ly$\beta$, and \ion{O}{6} $\lambda1035$.}\\
 & & & & & \\
\hline
\multirow{2}*{J1447} & \multirow{2}*{C, F, J, K, N, Q} & \multirow{2}*{$1084.8^{+8.9}_{-0.6}$} & \multirow{2}*{$-0.5 \pm 0.1$} & \multirow{2}*{$-2.2 \pm 0.2$} & \multirow{2}{0.33\textwidth}{NAL system at $z\approx1.03$: \ion{Mg}{2} $\lambda2800$, \ion{C}{4} $\lambda1549$, Lyman edge.\textsuperscript{b}}\\
 & & & & & \\
\hline
\multirow{2}*{J1530} & \multirow{2}*{B, G, H, L, P} & \multirow{2}*{$1090.5^{+12.9}_{-30.6}$} & \multirow{2}*{$-0.3^{+0.2}_{-0.1}$} & \multirow{2}*{$-2.0 \pm 0.2$} & \multirow{2}{0.33\textwidth}{\nodata}\\
 & & & & & \\
\hline
\multirow{2}*{J1629\textsuperscript{c}} & \multirow{2}*{E, H, L, O, R} & \multirow{2}*{\nodata} & \multirow{2}*{$+0.9^{+0.8}_{-1.4}$} & \multirow{2}*{$-1.6 \pm 0.1$} & \multirow{2}{0.33\textwidth}{HST spectrum appears reddened (see Appendix \ref{subsec:J1629}).}\\
 & & & & & \\
\enddata
\tablecomments{Column (1): object name. Column (2): rest-frame windows used in fitting a broken power law continuum to each object's full HST STIS spectrum (see Section \ref{subsec:cont}). Column (3): wavelength of the power law break. Column (4): best-fit power law index ($f_{\lambda} \propto \lambda^{\alpha_{\lambda}}$) and $1\sigma$ uncertainty for the broken power law fit to the HST FUV continuum (blueward of the break). Column (5): best-fit power law index and $1\sigma$ uncertainty for our broken power law fit to the HST NUV continuum (redward of the break), for all targets but J1629. Column (6): comments on identified NAL systems (see Section \ref{subsec:discuss_absorp}).\\
\textsuperscript{a}\,Continuum fit windows (in {\AA}) corresponding to letter indices in Column (2): A: 800$-$820. B: 810$-$820. C: 811$-$820. D: 850$-$880. E: 855$-$880. F: 850$-$870. G: 870$-$880. H: 1090$-$1105. I: 1085$-$1090. J: 1082$-$1088. K: 1097$-$1102. L: 1140$-$1155. M: 1140$-$1145. N: 1140$-$1149. O: 1280$-$1290. P: 1275$-$1284. Q: 1275$-$1285. R: 1315$-$1325. S: 1350$-$1362. T: 1440$-$1465.\\
\textsuperscript{b}\,Absorption system was tentatively identified by \citet{Seyffert13}.\\
\textsuperscript{c}\,The fit windows listed for SDSS J1629 are used only to find its FUV spectral index ($\alpha_{\rm FUV}$) from the HST spectrum (see Section \ref{subsec:cont}). Its NUV spectral index ($\alpha_{\rm NUV}$) is measured from the SDSS continuum, fit using the windows 2150$-$2250 and 3900$-$3940~{\AA}. We do not report a power law break value for this object.
}
\end{deluxetable*}

\subsection{Continuum} \label{subsec:cont}

To constrain the continuum shape of each HST UV spectrum, we fit a broken power law model of the form
\begin{displaymath}
f_{\lambda} \propto
\begin{cases}
\left( \lambda / \lambda_{\rm break} \right)^{\alpha_{\rm FUV}} & : \lambda < \lambda_{\rm break}\\
\left( \lambda / \lambda_{\rm break} \right)^{\alpha_{\rm NUV}} & : \lambda > \lambda_{\rm break}
\end{cases}
\end{displaymath}
where $\lambda_{\rm break}$ is the wavelength location of the break. Throughout the rest of this work, we define ``FUV" as the portion of the HST spectrum blueward of the break and ``NUV" as that redward of the break, so that the spectral indices of the corresponding continua are given by $\alpha_{\rm FUV}$ and $\alpha_{\rm NUV}$, respectively. We perform all spectral analyses in the rest-frame.

The exact range of HST rest-frame spectral coverage varies by target (illustrated in Figure \ref{fig:all_spectra}), with bounds falling between 675$-$842~{\AA} at the short-$\lambda$ end and 1310$-$1636~{\AA} at the long-$\lambda$ end. There are a number of corresponding rest-frame wavelength windows commonly used for continuum fitting throughout the literature \citep[e.g.,][]{Telfer0202,DS09,Stevans14}, chosen for their general dearth of emission line contamination. However, most of our six candidates show signatures of narrow absorption line (NAL) systems\footnote{Typically, narrow absorption lines (NALs) are defined as having FWHM $<500$ km s$^{-1}$, broad absorption lines (BALs) are defined as having FWHM $>2000$ km s$^{-1}$, and mini-BALs occupy the range between (see, e.g., \citealt{Weymann81}; \citealt{Hamann04} and references therein; \citealt{Gibson09}).} in their HST spectra, so we customized these windows as needed on a per-object basis. We provide the wavelength ranges adopted for each object in Table \ref{tab:windows}, along with basic 
information regarding absorption systems we identified via visual inspection of each spectrum.

We performed the model continuum fit to the observed flux densities (weighted by the $1\sigma$ uncertainty for each measurement) within these spectral windows via a $\chi^{2}$ minimization routine,\footnote{We used the \texttt{astropy.modeling} Python package to perform all fits \citep{astropy13, astropy18}.} and we imposed bounds on the break location of $950$~{\AA} $\leq \lambda_{\rm break} \leq 1300$~{\AA} \citep[based on, e.g.,][]{Zheng97,Telfer0202,Shang05,Stevans14}. The resulting best-fit spectral indices, $\alpha_{\rm FUV}$ and $\alpha_{\rm NUV}$, are given in Table \ref{tab:windows}. 

To estimate uncertainty in the spectral indices for each object, we used a Monte Carlo algorithm to generate a set of 5,000 mock spectra. We adjusted the flux density of each pixel within our fitting windows via random sampling of a normal distribution where the mean and standard deviation were set respectively to the observed flux density and uncertainty. We then re-fit a broken power law model to each simulated spectrum. We assigned $\pm1\sigma$ error bars by finding the 16\textsuperscript{th} and 84\textsuperscript{th} percentile values of the resulting distributions of $\alpha_{\rm FUV}$ and $\alpha_{\rm NUV}$ from each set of 5,000 simulated spectra. 

\begin{figure*}[tb!]
\plotone{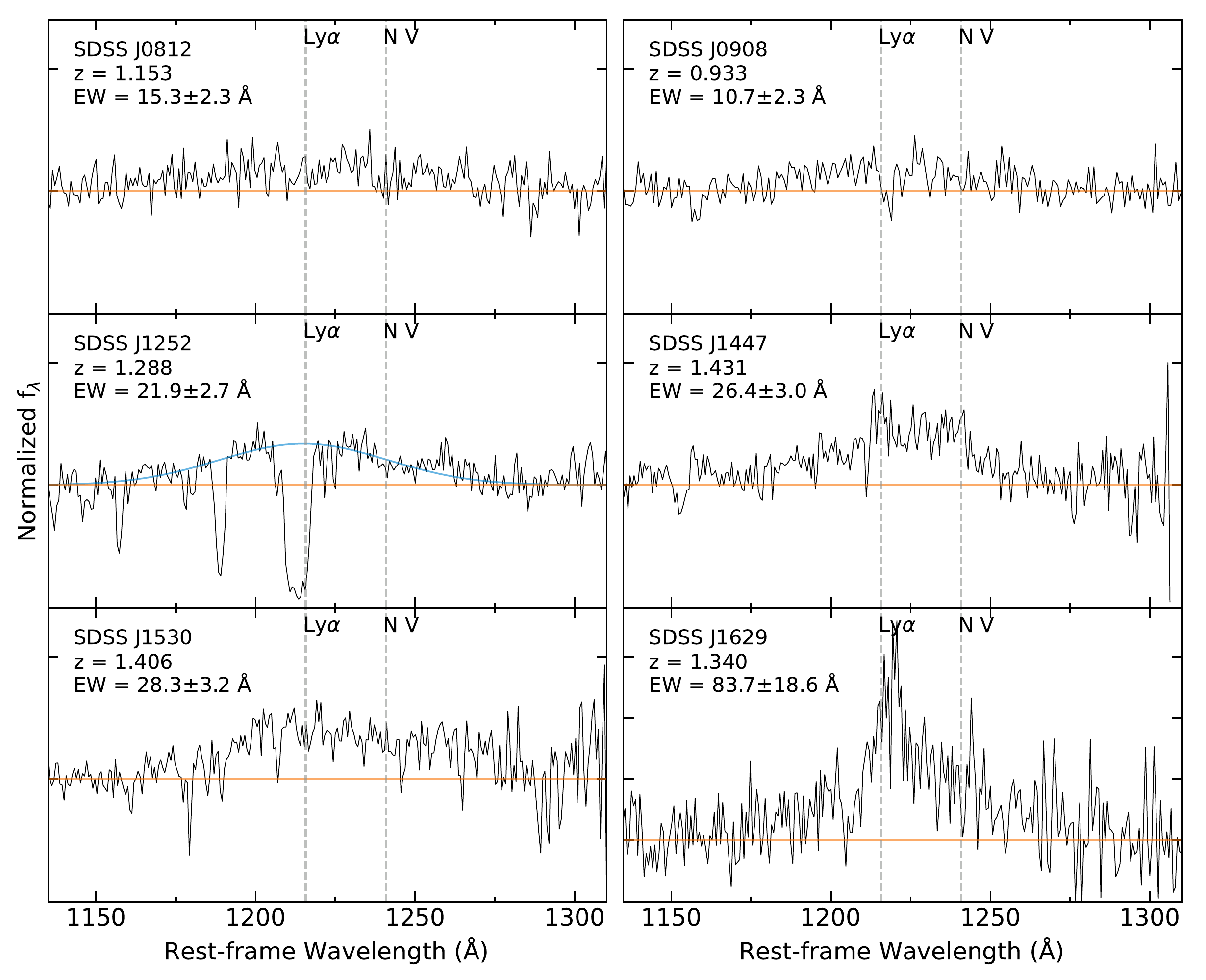}
\caption{Expanded view of the Ly$\alpha + $\ion{N}{5} complex for each WLQ candidate. The orange lines represent the best-fit linear continua. For SDSS J1252, the fitted Gaussian model of the emission complex is shown by a blue curve (see Section \ref{subsec:lyalpha}). Vertical dashed lines indicate the wavelength locations of Ly$\alpha$ and \ion{N}{5}. Note: all objects are shown with identical vertical scale except SDSS J1629, which is shown at a reduced scale to accommodate the full emission feature.
\label{fig:all_lyalpha}}
\end{figure*}

In principle, we could extend the wavelength coverage for our continuum fits by including emission-free regions from the SDSS spectrum of each target. However, there is a risk of flux variability between the SDSS and HST epochs, so we did not utilize the SDSS for our continuum fitting. The only exception is SDSS J1629, which shows a redder spectrum overall and a positive $\alpha_{\rm FUV}$. We attempted to fit a broken power law to only the HST spectrum, but extrapolating the NUV fit to longer wavelengths under-predicted the flux expected from the SDSS at an unreasonable level (see Figure \ref{fig:all_spectra}), almost certainly because the HST spectrum is more affected by absorption than the SDSS. Thus, for only this source, we fit the NUV continuum using the SDSS spectrum, adopting rest-frame windows 2150$-$2250 and 3900$-$3940~{\AA}. We include the resulting best-fit value as $\alpha_{\rm NUV}$ in Table \ref{tab:windows} for completeness, but because SDSS J1629 ultimately does not survive as a WLQ candidate (see Section \ref{sec:results}), the exact value adopted for its NUV continuum does not influence any final conclusions.

\subsection{Line Measurements} \label{subsec:lines}

To examine the properties of UV emission lines in our six candidate WLQs, we measured line strength via EW for the Ly$\alpha + $\ion{N}{5} complex. We also measured the EW and blueshift of broad \ion{C}{4} emission for SDSS J0908 (the only one of our six targets with this line within the spectral range of our HST observations). The spectral regions covering the Ly$\alpha + $\ion{N}{5} and \ion{C}{4} complexes are shown in Figures \ref{fig:all_lyalpha} and \ref{fig:c4_J0908}, respectively.

\subsubsection{Ly$\alpha + $\ion{N}{5}} \label{subsec:lyalpha}

We adopted the method described by \citetalias{DS09} (with some slight modifications) to measure the strength of the blended Ly$\alpha + $\ion{N}{5} complex. We fit a local linear continuum to each spectrum via a $\chi^{2}$ minimization routine using two 20~{\AA} windows centered on $\lambda$\textsubscript{rest} $=$ 1145~{\AA} and 1285~{\AA} (for SDSS J1252, we masked the ranges 1135$-$1138 and 1143$-$1150~{\AA} to avoid significant absorption features). We then performed a simple trapezoidal integration of the normalized flux above the best-fit linear continuum within the range 1160$-$1290~{\AA} to measure the rest-frame EW[Ly$\alpha$+\ion{N}{5}]. One candidate, SDSS J1252, shows strong absorption features over the broad Ly$\alpha$ emission (visible in Figure \ref{fig:all_lyalpha} and discussed further in Section \ref{subsec:discuss_absorp}); in this case only, we masked out the wavelengths affected by absorption, fit a single broad Gaussian to model the intrinsic emission profile, and measured EW[Ly$\alpha$+\ion{N}{5}] by integrating over the Gaussian model.\footnote{For the other five candidates, fitting a single broad Gaussian gave similar EW measurement results vs.\ performing numerical integration of the normalized flux (i.e., agreement on EW[Ly$\alpha$+\ion{N}{5}] within a few percent). Including multiple Gaussians did not improve the quality of the fit. Thus, for the other five candidates we prefer to use the numerical integration-based measurements for consistency with \citetalias{DS09}.}

\begin{deluxetable*}{lcccccc}[tb!]
\tablenum{3}
\tablecaption{Emission Line Measurements and Derived Quantities\label{tab:EWresults}}
\tablehead{
\colhead{Source Name} & \colhead{EW[Ly$\alpha$]} & \colhead{EW[\ion{C}{4}]} & \colhead{$\Delta v$[\ion{C}{4}]} & \colhead{EW[\ion{Mg}{2}]} & \colhead{$R_{\rm Ly\alpha,MgII}$} & \colhead{$M_{\rm NUV}$}\\
\colhead{} & \colhead{(\AA)} & \colhead{(\AA)} & \colhead{(km s$^{-1}$)} & \colhead{(\AA)} & \colhead{} & \colhead{(mag)}\\
\colhead{(1)} & \colhead{(2)} & \colhead{(3)} & \colhead{(4)} & \colhead{(5)} & \colhead{(6)} & \colhead{(7)}
}
\startdata
SDSS J0812 & $15.3 \pm 2.3$ & \nodata & \nodata & $8.4 \pm 0.7$ & $1.83 \pm 0.31$ & $-25.05 \pm 0.29$ \\
SDSS J0908 & $10.7 \pm 2.3$ & $8.3^{+3.8}_{-6.5}$ & $1741.9 \pm 249.6$ & $7.8 \pm 1.0$ & $1.37 \pm 0.34$ & $-24.21 \pm 0.57$ \\
SDSS J1252 & $21.9 \pm 2.7$ & \nodata & \nodata & $8.7 \pm 0.4$ & $2.51 \pm 0.33$ & $-25.60 \pm 0.31$ \\
SDSS J1447\textsuperscript{a} & $26.4 \pm 3.0$ & $7.7^{+0.2}_{-1.3}$ & $1319^{+759}_{-381}$ & $13.1^{+2.5}_{-0.1}$ & $2.06 \pm 0.45$ & $-24.89 \pm 0.57$\\
SDSS J1530 & $28.3 \pm 3.2$ & \nodata & \nodata & $12.9 \pm 0.4$ & $2.19 \pm 0.25$ & $-25.34 \pm 0.51$\\
SDSS J1629 & $83.7 \pm 18.6$ & \nodata & \nodata & $12.3 \pm 0.1$ & $6.81 \pm 1.70$ & $-25.62 \pm 0.30$\\
DS09 quasars\textsuperscript{b} & $63.6^{+38.3}_{-24.0}$ & $41.9^{+25.5}_{-15.8}$ & \nodata & \nodata & \nodata & \nodata\\
Int-$z$ quasars\textsuperscript{c} & \nodata & $36.6^{+24.9}_{-14.8}$ & \nodata & \nodata & \nodata & \nodata\\
Low-$z$ quasars\textsuperscript{d} & \nodata & \nodata & \nodata & $30.7^{+10.9}_{-8.0}$ & \nodata & \nodata\\
VB01 composite\textsuperscript{e} & $92.91 \pm 0.72$ & $23.78 \pm 0.10$ & \nodata & $32.28 \pm 0.07$ & $2.88 \pm 0.02$ & \nodata\\
\enddata
\tablecomments{Column (1): object name. Column (2): measured rest-frame EW and approximate $\pm1\sigma$ uncertainty for the Ly$\alpha + $\ion{N}{5} blend (see Section \ref{subsec:lyalpha}). Columns (3): measured rest-frame EW and approximate $\pm1\sigma$ uncertainty for \ion{C}{4} emission (see Section \ref{subsec:c4}). Column (4): line-of-sight blueshift (defined to be positive for approaching motion) and approximate $\pm1\sigma$ uncertainty for \ion{C}{4} emission. Column (5): Rest-frame EW and $1\sigma$ uncertainty for \ion{Mg}{2} emission, obtained from \citet{Shen11} except when noted otherwise. Column (6): ratio of Ly$\alpha$ to \ion{Mg}{2} line strength, $R_{\rm Ly\alpha,MgII} =$ EW[Ly$\alpha$+\ion{N}{5}]$/$EW[\ion{Mg}{2}] (see Section \ref{subsec:discuss_R}). Column (7): K-corrected NUV absolute magnitude (see Section \ref{subsec:results_absmag}).\\
\textsuperscript{a}\ion{C}{4} and \ion{Mg}{2} measurements for SDSS J1447 are obtained from \citetalias{Plotkin15}.\\
\textsuperscript{b}For convenience, the final four rows provide quantities from various comparison samples (see Section \ref{subsec:comp_sample} for further description). This row gives the mean values of the log-normal distributions of EW[Ly$\alpha$+\ion{N}{5}] and EW[\ion{C}{4}] in the \citetalias{DS09} quasar sample. The quoted uncertainties are the $1\sigma$ ranges of each distribution.\\
\textsuperscript{c}This row gives the mean value of the log-normal distribution of EW[\ion{C}{4}] in the ``intermediate-redshift" quasar sample (see Section \ref{subsec:comp_sample}). The quoted uncertainty is the $1\sigma$ range of the distribution.\\
\textsuperscript{d}This row gives the mean value of the log-normal distribution of EW[\ion{Mg}{2}] in the ``low-redshift" quasar sample (see Section \ref{subsec:comp_sample}). The quoted uncertainty is the $1\sigma$ range of the distribution.\\
\textsuperscript{e}This row quotes emission line measurements from the \citet{VB01} composite quasar spectrum.
}
\end{deluxetable*}

Uncertainties on the best-fit line measurements are dominated by systemic uncertainty in our ability to accurately place the local continuum level. To account for both statistical and systemic uncertainties, we adopted the scheme described in Appendix B.1 of \citetalias{Plotkin15}. We began with the best-fit local linear continuum and recorded both goodness of fit $\chi^{2}_{\rm best}$ and degrees of freedom $\nu_{\rm best}$. We then generated a $51\times51$ grid of mock linear continua for each spectrum by varying the observed flux density within each continuum fit window by evenly-stepped factors (ranging between approximately 0.8 and 1.2, depending on the source) across the grid and re-fitting the linear continuum. We compared each $i^{th}$ mock continuum model in the grid to the original spectrum, recording $\chi^{2}_{i}$, computing the relative $\Delta\chi^{2}_{i}=\chi^{2}_{i}-\chi^{2}_{\rm best}$, and measuring EW$_{i}$. The $68\%$ ($\sim$1$\sigma$) confidence interval corresponds to continua with $\Delta\chi^{2}_{i}=2.3$ \citep{Avni76}. From all continua with $|\Delta\chi^{2}_{i}|$ within this $1\sigma$ confidence range, we recorded the corresponding maximum and minimum EW$_{i}$ values, and compared these to the best-fit EW measurements to determine approximate $\pm1\sigma$ uncertainties. EW[Ly$\alpha$+\ion{N}{5}] best-fit measurements and uncertainties for all six candidates are given in Table \ref{tab:EWresults}.

\subsubsection{\ion{C}{4}} \label{subsec:c4}

\begin{figure}[tb!]
\plotone{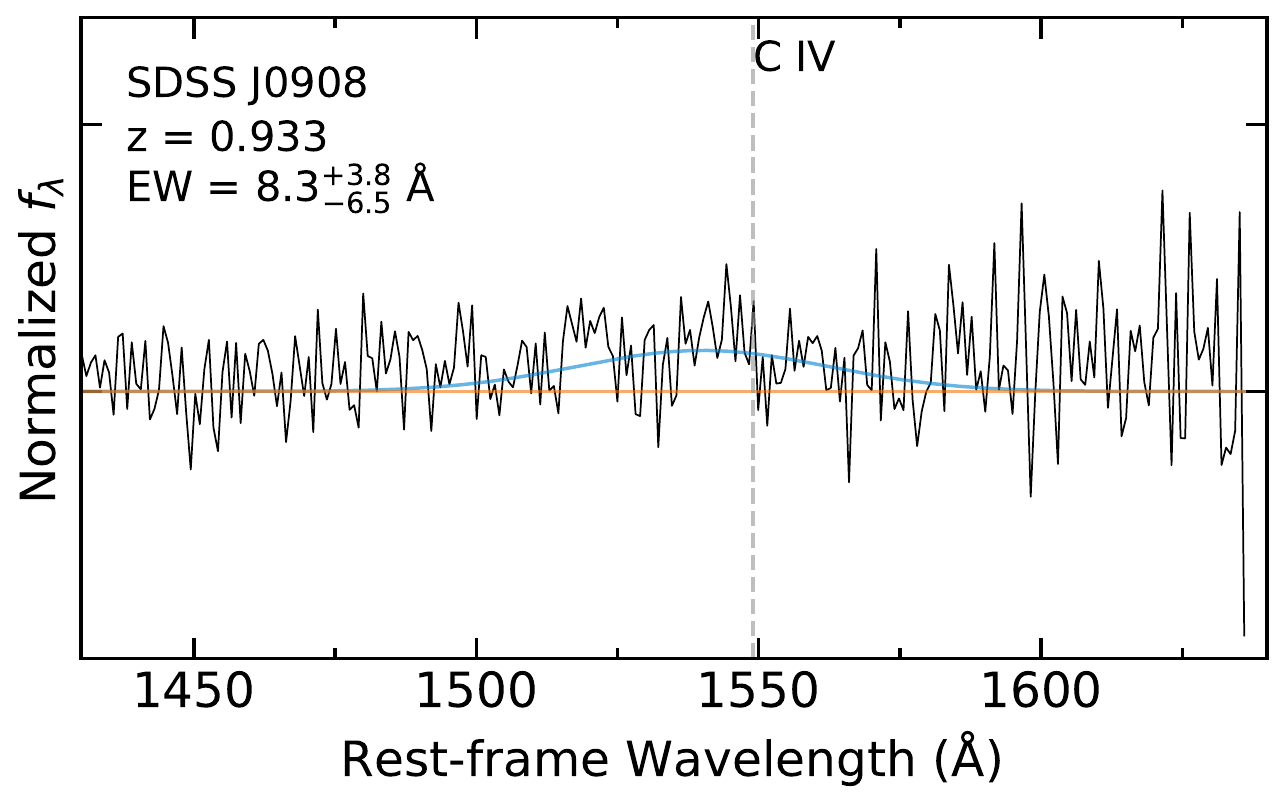}
\caption{Expanded view of the \ion{C}{4} complex for SDSS J0908. The orange line represents the best-fit linear continuum. The Gaussian model fitted to the emission complex is shown by a blue curve (see Section \ref{subsec:c4}). The vertical dashed line gives the \ion{C}{4} rest-frame wavelength location to illustrate the blueshift of the observed emission feature.
\label{fig:c4_J0908}}
\end{figure}

The HST spectral coverage for SDSS J0908 runs redward to $\lambda$\textsubscript{rest} $\approx 1630$~{\AA}, making it unique among our observations in that it provides coverage of \ion{C}{4} emission. For this target, we measured EW[\ion{C}{4}] and its uncertainty in a manner similar to that described in Section \ref{subsec:lyalpha}, with the local linear continuum fit using windows at $\lambda$\textsubscript{rest} $=$ 1435$-$1455 and 1570$-$1580~{\AA}, a single Gaussian curve fit to the emission line feature (shown in Figure \ref{fig:c4_J0908}), and integration performed over the range 1450$-$1580~{\AA}.\footnote{Measuring EW[\ion{C}{4}] via numerical integration of the normalized flux vs.\ fitting a single Gaussian provided similar results. Including multiple Gaussians did not improve the quality of the fit. We use the measurement from the Gaussian fit for consistency with \citetalias{DS09}, who adopted EW[\ion{C}{4}] measurements derived from Gaussian emission-line fits performed by \citet{Shen08}.} We also measured the blueshift of the emission line from 1549~{\AA} to the centroid of the Gaussian fit. The results are provided in Table \ref{tab:EWresults}.

\subsection{Luminosity-Matched Samples of Comparison Radio-Quiet Quasars} \label{subsec:comp_sample}

To derive a sample of ``normal" radio-quiet quasars for comparison of optical$-$UV properties, we considered objects from the SDSS DR7 quasar sample \citep[][]{Schneider10} and the \citet{Shen11} catalog. We limited the selection to quasars targeted for spectroscopy using the algorithm of \citet{Richards02}, and excluded objects identified by \citet{Shen11} as BAL or radio-loud quasars. We also excluded any objects identified by \citet{Collinge05} or \citet{Plotkin10b} as ``weak-featured." To ensure we compare our WLQ candidates to radio-quiet quasars of similar optical$-$UV luminosity and redshift, we limited our selection to the ranges $30.5 \leq \log l$\textsubscript{2500 {\AA}} $\leq 31.5$ (erg s$^{-1}$ Hz$^{-1}$) and $0.9 < z < 2.5$. We then found available $m_{\rm NUV}$ measurements by correlating the remaining objects to the GALEX UV catalogs \citep{Bianchi17} with a $3''$ match radius, excluding any match with a poor quality (i.e., artifact or extractor) flag. There are 14,713 radio-quiet quasars in the resulting set, which we partition further by designating a ``low-redshift" quasar sample with $0.9 < z < 1.5$ (6,381 quasars) and an ``intermediate-redshift" quasar sample with $1.5 < z < 2.5$ (8,332 quasars). Any potentially unidentified WLQs remaining in the sample are likely too few in number to have a significant impact on the statistics.

For easy comparison with average quantities, we add the following information to Table \ref{tab:EWresults} beneath our six targets. We list the mean values of the log-normal distributions of EW[Ly$\alpha$+\ion{N}{5}] and EW[\ion{C}{4}] in the high-redshift ($z>3$) quasar sample of \citetalias{DS09}, the mean of the log-normal distribution of EW[\ion{C}{4}] in the ``intermediate-redshift" quasar sample (using measurements from \citealt{Shen11}, limited to 7,906 objects with $>250$ good pixels for \ion{C}{4}), and the mean of the log-normal distribution of EW[\ion{Mg}{2}] (again using measurements from \citealt{Shen11}, limited to 6,341 objects with $>250$ good pixels for \ion{Mg}{2}) in the ``low-redshift" quasar sample.\footnote{We find that the distributions of EW measurements from our comparison quasar sample are well-described by a log-normal function. Note also that our choice of sample for each emission line is dictated primarily by the redshift range at which the line appears in the SDSS spectral window.} We adopt the $1\sigma$ range of each distribution as the quoted uncertainty. In the final row of Table \ref{tab:EWresults}, we quote Ly$\alpha$, \ion{C}{4}, and \ion{Mg}{2} EW measurements, as well as the ratio $R_{\rm Ly\alpha,MgII} =$ EW[Ly$\alpha$+\ion{N}{5}]$/$EW[\ion{Mg}{2}] (discussed in Section \ref{subsec:discuss_R}), from the \citet{VB01} composite quasar.

\section{Results} \label{sec:results}

\begin{figure}[tb!]
\plotone{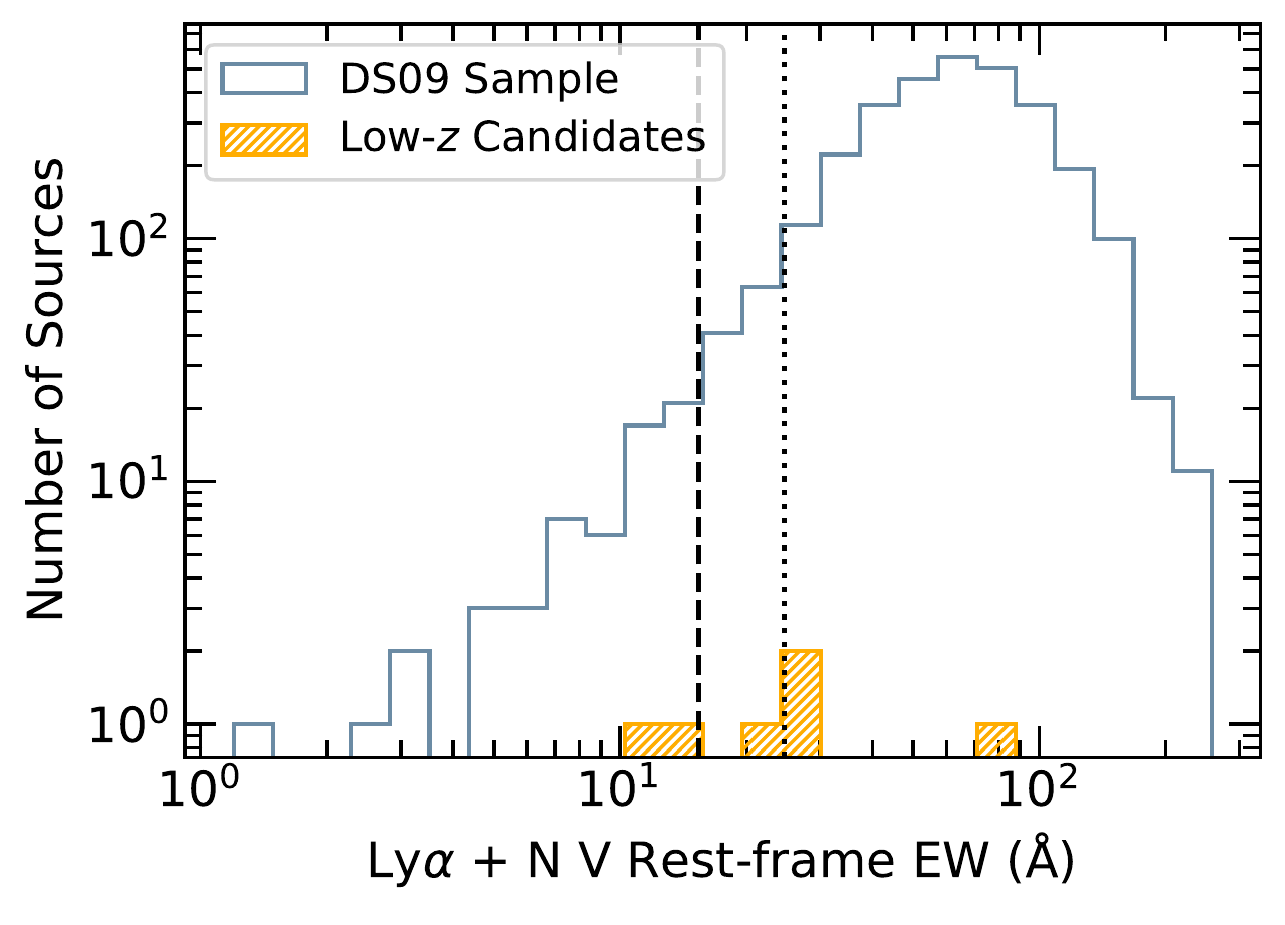}
\caption{Histogram of the EW[Ly$\alpha$+\ion{N}{5}] distribution for our candidate WLQs (orange bins) vs.\ the full \citetalias{DS09} catalog (blue-outlined bins). Both axes are given on a log scale. The vertical dashed line shows the $3\sigma$ ($<15.4$~{\AA}) weak threshold for WLQ classification, and the vertical dotted line shows the $2\sigma$ ($<27.4$~{\AA}) weak threshold. Two of our targets (SDSS J0812 and SDSS J0908) meet the \citetalias{DS09} criterion for WLQ classification, while one (SDSS J1629) is typical of normal quasars, and the remaining three are somewhat weak ($\sim2\sigma$) compared to the mean of the best-fit log-normal distribution (see \citetalias{DS09}). \label{fig:EW_hist}}
\end{figure}

Figure \ref{fig:EW_hist} shows the distribution of EW[Ly$\alpha$+\ion{N}{5}] values for our targets in comparison to the \citetalias{DS09} WLQ population. According to the \citetalias{DS09} definition (EW[Ly$\alpha$+\ion{N}{5}] $<$ 15.4~{\AA}), only two of our targets qualify as WLQs: SDSS J0812 and SDSS J0908, with EW[Ly$\alpha$+\ion{N}{5}] $= 15.3 \pm 2.3$ and $10.7 \pm 2.3$~{\AA}, respectively. SDSS J0908 further qualifies as a WLQ according to its weak EW[\ion{C}{4}] $= 8.3^{+3.8}_{-6.5}$~{\AA} (see Section \ref{sec:intro}).

Of the remaining four objects, three have EW[Ly$\alpha$+\ion{N}{5}] $\approx 20-30$~{\AA}, which is still relatively weak compared to the \citet{VB01} composite quasar spectrum (EW[Ly$\alpha$] $= 92.91 \pm 0.72$~{\AA}). For reference, the $>2\sigma$-weak tail of the \citetalias{DS09} WLQ distribution corresponds to EW[Ly$\alpha$+\ion{N}{5}] $<24.7$~{\AA}. Considering that WLQs represent one extreme of a continuous population, it is therefore likely that these three objects, while not \textit{bona fide} WLQs, still represent the intermediate part of the population that connects WLQs to normal radio-quiet quasars (as explored for \ion{C}{4} by \citealt{Ni18}). This matter introduces an important caveat; while we adopt the same $3\sigma$ and $2\sigma$ weakness thresholds in order to classify our targets in relation to the known population, \citetalias{DS09} (see Section 2 therein) stress that these boundaries are motivated primarily by the statistics of their sample and are therefore somewhat arbitrary. Until a more robust, physically-motivated set of criteria for WLQ selection can be established, the accuracy and precision of such thresholds are uncertain.

The final target, SDSS J1629, has a highly-reddened HST spectrum and shows EW[Ly$\alpha$+\ion{N}{5}] $= 83.7 \pm 18.6$~{\AA} (i.e., typical of normal radio-quiet quasars). This source clearly does not survive as a WLQ candidate, and it is excluded from the remaining discussion (for completeness, its individual properties are discussed in Appendix \ref{subsec:J1629}).

\subsection{UV Continuum Properties} \label{subsec:results_absmag}

Our measured NUV spectral indices ($\alpha_{\rm NUV}$; Table \ref{tab:windows}) are generally typical of normal quasars (compared to, e.g., $\alpha_{\rm NUV} = -1.54$ from \citealt{VB01} and the $\alpha_{\rm NUV}$ distribution shown in Figure 2 of \citealt{Ivashchenko14}). Except for SDSS J1629, our measured $\alpha_{\rm FUV}$ values are also generally consistent with normal quasars (e.g., $\alpha_{\rm FUV} = -0.43$ for radio-quiet quasars from \citealt{Telfer0202}).

Further supporting typical continuum shapes, the observed-frame NUV luminosities of our targets are similar to normal radio-quiet quasars at comparable redshifts and observed-frame optical luminosities. In Table \ref{tab:EWresults} we tabulate $K$-corrected (to $z=0$) absolute NUV magnitudes for our targets, $M_{\rm NUV} = m_{\rm NUV} - DM - K$, where $m_{\rm NUV}$ is the GALEX NUV apparent magnitude \citep{Bianchi17}, $DM$ is the distance modulus, and $K=2.5\,(1+\alpha_{\rm NUV}) \log(1+z)$ is the $K$-correction (see, e.g., \citealt{Hogg99}) for which we adopt the best-fit $\alpha_{\rm NUV}$ from each of our candidates. Our targets have a mean absolute magnitude $\langle M_{\rm NUV} \rangle = -25.02 \pm 0.47$ (the quoted error is the standard deviation). 

We compare our targets' $M_{\rm NUV}$ values to those of the ``low-redshift" quasar sample found in Section \ref{subsec:comp_sample} (calculated as described above, but assuming $\alpha_{\rm NUV} =-1.54$). For this sample, $\langle M_{\rm NUV} \rangle = -24.61 \pm 0.65$ (the quoted error is the standard deviation). Given the small number of HST targets, we employ three different non-parametric statistical tests to compare the distributions of $M_{\rm NUV}$, including K-S and A-D tests (both of which compare the distributions of two different populations) and a Mann-Whitney (M-W) test (which compares the means of two populations). No test rejects the null hypothesis, providing $p=0.13$, 0.17, and 0.06 for the K-S, A-D, and M-W tests, respectively. Thus, the UV luminosities of our WLQ candidates appear to be consistent with those of other SDSS-selected radio-quiet quasars at similar redshifts and with similar optical luminosities, with the caveat of small number statistics.

\subsection{Comparison of Ly$\alpha$ with X-ray Properties}
\label{subsec:results_x-ray}

\begin{figure*}[tb!]
\gridline{\fig{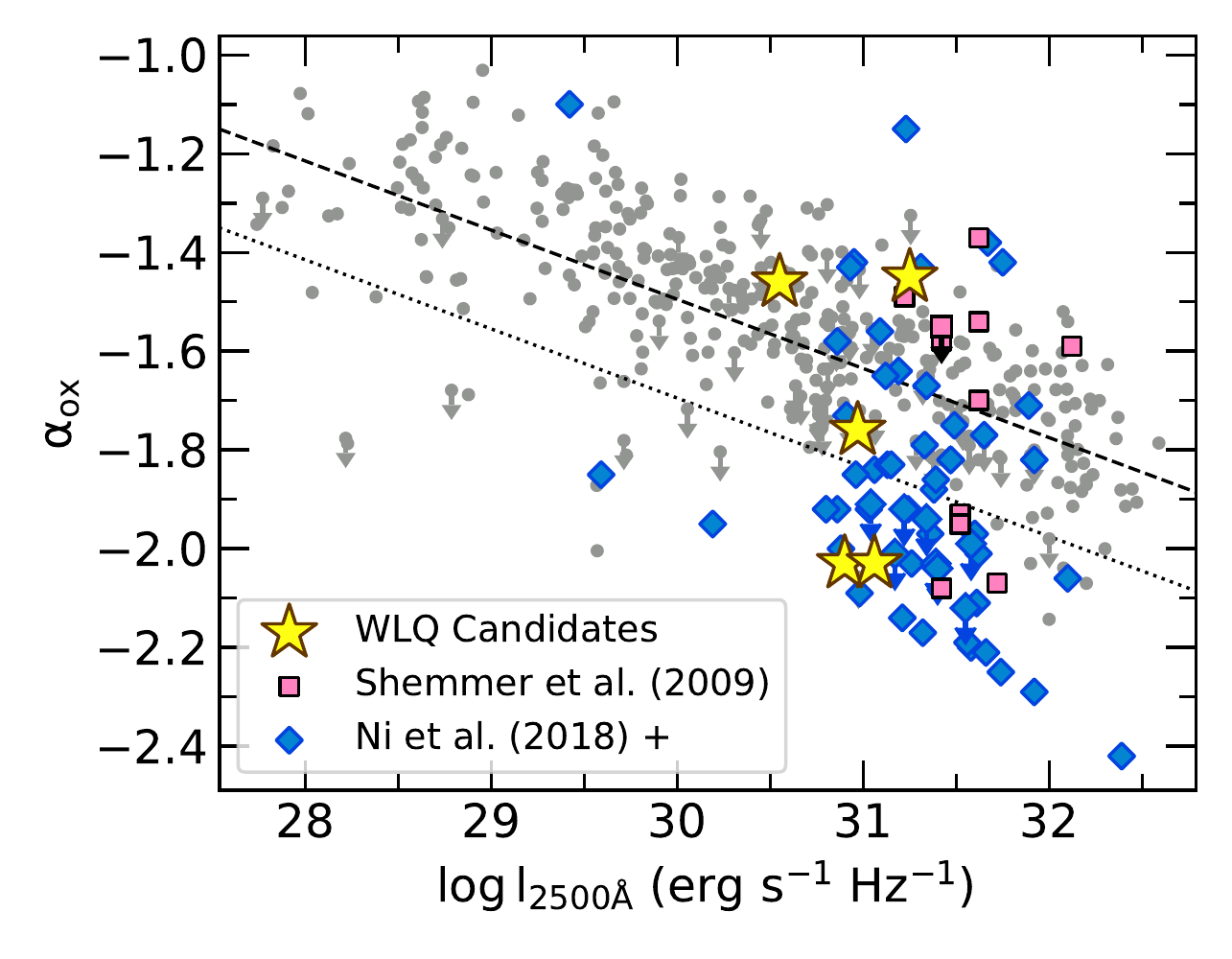}{0.45\textwidth}{(a)}
          \fig{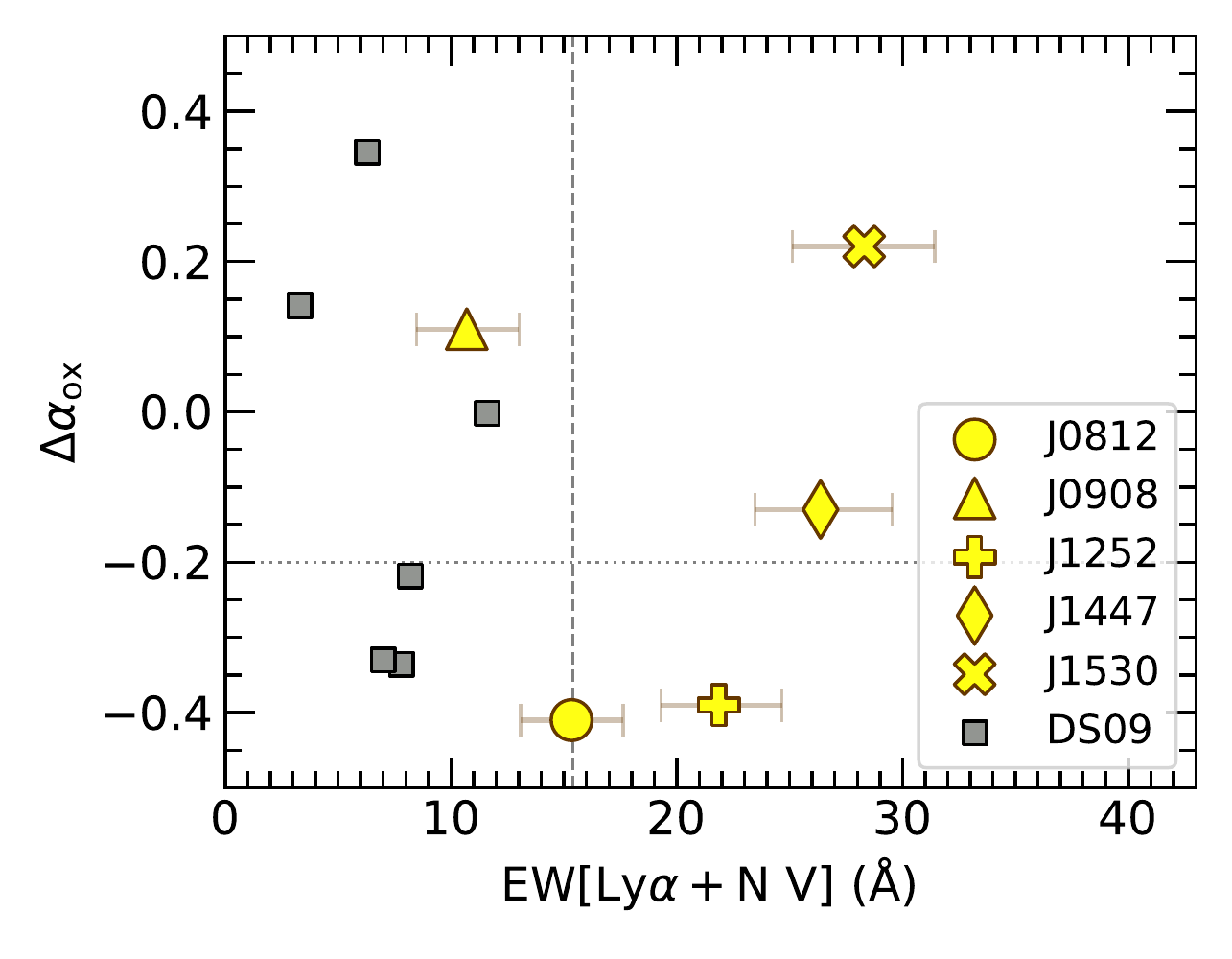}{0.45\textwidth}{(b)}}
\caption{(a) X-ray to optical flux ratio $\alpha_{\rm ox}$ vs.\ specific luminosity at 2500~{\AA}. The HST targets are shown by yellow-filled stars, the WLQ sample of \citet{Shemmer09} by pink-filled squares, and the \citet{Ni18} extreme and intermediate WLQ samples (including objects from \citealt{Wu12} and \citealt{Luo15}) by blue-filled diamonds. Grey dots represent the combined \citet{Steffen06} and \citet{Just07} parent quasar samples. Downward arrows represent upper limits on $\alpha_{\rm ox}$ from X-ray non-detections. We adopt $l$\textsubscript{2500 {\AA}} directly from each reference because the scale shown is not visibly sensitive to minor changes in the adopted cosmology. The dashed line indicates the $\alpha_{\rm ox}$ and $l$\textsubscript{2500 {\AA}} anticorrelation slope given by Eq. (3) of \citet{Just07}, while the dotted line shows the conventional demarcation between X-ray normal and X-ray weak, $\Delta\alpha_{\rm ox}\leq-0.2$. Our targets have luminosities roughly similar to each other, and they span the range of $\Delta\alpha_{\rm ox}$ for WLQs. (b) $\Delta\alpha_{\rm ox}$ vs.\ EW[Ly$\alpha$+\ion{N}{5}]. Yellow-filled symbols are our HST targets (SDSS designations are given in the legend). The grey-filled squares denote \citetalias{DS09} WLQs for which we have found $\Delta\alpha_{\rm ox}$ using $f_{\rm 2 keV}$ measurements from \citet{Shemmer09} and $f$\textsubscript{2500 {\AA}} from \citet{Shen11}. The horizontal dotted line indicates the $\Delta\alpha_{\rm ox}\leq-0.2$ demarcation for X-ray weakness, while the vertical dashed line is the EW[Ly$\alpha$+\ion{N}{5}] $< 15.4$~{\AA} WLQ cutoff. For these samples (both combined and separately), we do not have a correlation between X-ray weakness and Ly$\alpha + $\ion{N}{5} emission (see Section \ref{subsec:results_x-ray}). \label{fig:aox_L2500}}
\end{figure*}

To examine our targets in the context of the varied X-ray properties displayed by WLQs (see Section \ref{sec:intro}), we compare their Ly$\alpha$ and observed X-ray emission strengths. All of our HST targets have been observed by the Chandra X-ray Observatory, and we adopt the $f_{\rm 2 keV}$ flux densities tabulated by \citet{Wu12} and \citet{Luo15} in addition to the $f$\textsubscript{2500 {\AA}} flux densities from \citet{Shen11}. Following \citet{Tananbaum79}, we define an object's optical$-$UV to X-ray spectral slope as 
\begin{displaymath}
\alpha_{\rm ox} = 0.3838\log(f_{\rm 2 keV} / f\textsubscript{2500 {\AA}}).
\end{displaymath}
We also adopt the X-ray weakness parameter, 
\begin{displaymath}
\Delta \alpha_{\rm ox} = \alpha_{\rm ox} - \alpha_{\rm ox,qso},
\end{displaymath}
where $\alpha_{\rm ox,qso}$ is the value predicted by the $\alpha_{\rm ox}$ and $l\textsubscript{2500 {\AA}}$ anticorrelation displayed by broad-line quasars (for consistency with prior WLQ studies, we adopt the best-fit relationship given by Eq. (3) of \citealt{Just07}; see also, e.g., \citealt{Steffen06,Lusso10,Timlin20}).\footnote{\citet{Timlin20} suggest an intrinsic scatter of $\pm 0.11$~dex for $\alpha_{\rm ox,qso}$ and note that within their range of uncertainty, their best-fit relation is consistent with that of  \citet{Just07}.} Our targets' $\alpha_{\rm ox}$ and $\Delta\alpha_{\rm ox}$ values are given in Table \ref{tab:obslog}, and Figure \ref{fig:aox_L2500}a shows them in relation to the WLQ samples of \citet{Shemmer09}, \citet{Wu12}, \citet{Luo15}, and \citet{Ni18}, as well as quasars from \citet{Steffen06} and \citet{Just07}.

We compare $\Delta\alpha_{\rm ox}$ against EW[Ly$\alpha$+\ion{N}{5}] for our targets combined with the small subset of 6 \citetalias{DS09} WLQs possesssing corresponding $f_{\rm 2 keV}$ measurements in the literature (obtained from \citealt{Shemmer09}),\footnote{This comparison sample is higher-redshift ($z>3$) and slightly higher in UV specific luminosity ($\log l$\textsubscript{2500 {\AA}} $\sim 31.5$ erg s$^{-1}$ Hz$^{-1}$; see the pink-filled squares in Figure \ref{fig:aox_L2500}a) than our targets. While not ideal, it currently represents our best means for comparing the combination of $\Delta\alpha_{\rm ox}$ and EW[Ly$\alpha$+\ion{N}{5}] between our HST targets and other WLQs.} shown in Figure \ref{fig:aox_L2500}b. Within this sample, there is no correlation between X-ray weakness and Ly$\alpha$ weakness: a Pearson correlation test gives a correlation coefficient $r=0.17$ ($p=0.78$) for our 5-target subset alone, and $r=-0.07$ ($p=0.84$) for the $N=11$ combined sample. This result is consistent with the results of \citet{Ni18,Ni22} and \citet{Timlin20} for \ion{C}{4} in WLQs.

\section{Discussion} \label{sec:discussion}

We obtained UV spectra of six candidate low-redshift ($0.9 < z < 1.5$) WLQs using STIS on HST. Our targets were selected primarily on the basis of weak \ion{Mg}{2} emission in their SDSS spectra (described in Section \ref{subsec:samp} and tabulated in Table \ref{tab:EWresults}), and we have extended their UV coverage blueward to $\sim$700$-$800~{\AA} in the rest-frame. In addition to covering the Ly$\alpha$ + \ion{N}{5} complex, this expanded spectral coverage allows more complete constraints on the UV continuum, as well as comparison of the relative strengths of high- and low-ionization emission lines. As noted in Section \ref{sec:results}, two of our targets qualify as WLQs based on the weakness of their Ly$\alpha$ + \ion{N}{5} emission, and another three may represent the intermediate part of the population (hereafter, simply ``intermediate quasars") between WLQs and typical radio-quiet quasars. We begin this section with a brief discussion of SDSS J0908, which represents our best example of a low-redshift WLQ. More detailed information on each individual target, including SDSS J0908, is provided in Appendix \ref{sec:app_indiv}.

\subsection{SDSS J0908, a Low-$z$ WLQ}
\label{subsec:discuss_J0908}

SDSS J0908 has the weakest Ly$\alpha$ + \ion{N}{5} emission among our HST targets, and it also survives as a WLQ on account of its weak broad \ion{C}{4} emission. We quantify the weakness of its \ion{C}{4} emission in the context of the Modified Baldwin Effect (e.g., \citealt{Baldwin77,Baskin04,Shemmer15}), whereby in normal radio-quiet quasars, EW[\ion{C}{4}] is observed to decrease with increasing Eddington ratio ($L_{\rm bol}/L_{\rm Edd}$, a parameterization of the mass-weighted accretion rate). It appears that WLQs tend not to follow this anticorrelation, instead having EW values significantly lower than expected for their $L_{\rm bol}/L_{\rm Edd}$ (\citealt{Shemmer15}). In SDSS J0908, EW[\ion{C}{4}] is a factor of $\sim$2 weaker than expected (at a significance $>3\sigma$) for its $L_{\rm bol}/L_{\rm Edd}$.\footnote{As found via the difference between the measured value and the best-fit relation given by Eq. (2) of \citet{Shemmer15}, using $L_{\rm bol}/L_{\rm Edd}$ from \citet{Shen11}.}

SDSS J0908 also shows moderate blueshift of the broad \ion{C}{4} emission line ($\Delta v$[\ion{C}{4}] $= 1741.9 \pm 249.6$ km s$^{-1}$ relative to the systemic redshift). We interpret this blueshift in the context of ``disk-wind" models \citep[e.g.,][]{Emmering92,Murray95,Marziani96,Elvis00,Leighly04,Richards11}, in which high-ionization BELR emission is dominated at one end of observed EW$-$blueshift parameter space by a disk (or failed wind) component with high EW and low blueshift, and at the other end by a wind component with low EW and high blueshift, likely driven by increasing $L_{\rm bol}/L_{\rm Edd}$ \citep{Giustini19}. Most relevant here is that some WLQs occupy a distinctly wind-dominated regime (e.g., \citealt{Wu12}; \citealt{Luo15}; \citetalias{Plotkin15}). The combination of moderately weak and blueshifted \ion{C}{4} and only mildly weak \ion{Mg}{2} (see Table \ref{tab:EWresults}) in SDSS J0908 is consistent with the disk-wind model and the observed trend for WLQs to move into the wind-dominated regime as \ion{C}{4} EW diminishes (for visual comparison, see Figure 8 of \citetalias{Plotkin15}).

These results present a curious contrast between the properties of SDSS J0908 and another of our targets, SDSS J1447, for which \ion{C}{4} was measured by \citetalias{Plotkin15}. As noted above, EW[\ion{C}{4}] in SDSS J0908 is more than $3\sigma$ weaker than expected from the Modified Baldwin Effect. On the other hand, SDSS J1447, despite being weak in \ion{C}{4} per the \citetalias{DS09} paradigm, was found by \citetalias{Plotkin15} to be only $\sim$1.5$\sigma$ weaker than expected from the Modified Baldwin Effect (although see caveat below regarding $L_{\rm bol}/L_{\rm Edd}$ derived from virial black-hole mass estimates). Accordingly, \citetalias{Plotkin15} predicted that SDSS J1447 is not a WLQ. Indeed, we find that it is insufficiently weak in Ly$\alpha$ + \ion{N}{5} emission to qualify as a WLQ (although we reiterate that it may be an intermediate quasar), and its ratio of EW[Ly$\alpha$+\ion{N}{5}] to EW[\ion{C}{4}] is higher than SDSS J0908. \citet{Dietrich02} suggested that the (classical, luminosity-based) Baldwin Effect for high-ionization lines in typical radio-quiet quasars depends partly on the ionization energy of the species, and that the anticorrelation slope is steeper for \ion{C}{4} than for Ly$\alpha$. If this suggestion holds true for the \textit{Modified} Baldwin Effect, we speculate that the ratio EW[Ly$\alpha$+\ion{N}{5}]$/$EW[\ion{C}{4}] in normal radio-quiet quasars might increase with $L_{\rm bol}/L_{\rm Edd}$, and weak \ion{C}{4} in very luminous quasars may not always guarantee weak Ly$\alpha$. Therefore, when attempting to classify high-$L_{\rm bol}/L_{\rm Edd}$ WLQs in the absence of Ly$\alpha$ coverage, it may be useful to first verify that EW[\ion{C}{4}] departs significantly from the Modified Baldwin Effect. However, we caution that the above comparison and interpretation may suffer from substantial systematic uncertainties associated with single-epoch virial black-hole mass estimation techniques \citep[e.g.,][and references therein]{Shen13}. The \citetalias{Plotkin15} estimate for SDSS J1447 is H$\beta$-based, while the \citet{Shen11} estimates for our remaining targets (including SDSS J0908) are \ion{Mg}{2}-based and are likely to be affected by non-virial motions of the broad line gas (e.g., \citealt{Wu11}; \citetalias{Plotkin15}; \citealt[][submitted]{Yi22}).

\citet[][submitted]{Rivera22} suggest parameterizing quasars using the ``\ion{C}{4} distance," as it evaluates objects in relation to a best-fit polynomial curve in EW$-$blueshift parameter space (e.g., see Fig. 12 of \citealt{Rivera20}) and may provide a better indicator of $L_{\rm bol}/L_{\rm Edd}$ than the Modified Baldwin Effect. Intriguingly, even in ``\ion{C}{4} distance" space, some WLQs could still be outliers. For example, SDSS J0908 and J1447 occupy similar positions in EW$-$blueshift parameter space (and thus have similar \ion{C}{4} distances), yet they display different Ly$\alpha$ properties as discussed in the previous paragraph (with the above caveat of black-hole mass estimation methods).

\subsection{Comparison of Optical$-$UV and X-ray Properties} \label{subsec:discuss_R}

The X-ray and multiwavelength properties of our targets were previously examined by \citet{Wu12} and \citet{Luo15}, and were found to be most consistent with the slim-disk scenario (Section \ref{sec:intro}). The expanded wavelength coverage afforded by the HST spectra complements these studies by allowing further tests of the shapes of the objects' ionizing continua. Notably, we have found a lack of correlation between EW[Ly$\alpha$+\ion{N}{5}] and X-ray weakness (Section \ref{subsec:results_x-ray}) that is also consistent with the slim-disk scenario, as it implies the BELR is exposed to an X-ray$-$optical SED different from the one we observe.

We are also now able to consider the ratio $R_{\rm Ly\alpha,MgII} =$ EW[Ly$\alpha$+\ion{N}{5}]$/$EW[\ion{Mg}{2}] (see Table \ref{tab:EWresults}; except where noted, \ion{Mg}{2} measurements are taken from \citealt{Shen11}). Unfortunately, we are not aware of a sizable sample of quasars in our targets' redshift range with uniform coverage of both the Ly$\alpha$ and \ion{Mg}{2} complexes, so we limit our comparison to the ratio of ``typical" EW values from the \citet{VB01} composite quasar spectrum (listed in Table \ref{tab:EWresults}), where $R_{\rm Ly\alpha,MgII} = 2.88$. All of our targets fall below this ratio, with a mean of $\langle R_{\rm Ly\alpha,MgII} \rangle = 2.0 \pm 0.4$ (the quoted error is the standard deviation). The lowest value belongs to our best-confirmed WLQ, SDSS J0908, and for objects with larger EW[\ion{Mg}{2}], the ratio tends toward the expectation for typical quasars. This result implies that while both species are weaker than normal, they do not diminish equally; Ly$\alpha$ is \textit{preferentially} weaker than \ion{Mg}{2}, which is generally expected from a soft ionizing SED (and disfavors previous suggestions that WLQ BELRs could be gas deficient). This result is also consistent with studies of other high- to low-ionization line ratios, such as \ion{C}{4} to \ion{Mg}{2} and H$\beta$ (e.g., \citealt{Wu12}; \citetalias{Plotkin15}), and it highlights limitations of basing searches for low-$z$ WLQs solely on the properties of low-ionization lines.

Given the slim-disk model's emphasis on super-Eddington accretion, it is intriguing that our bona fide WLQs do not have particularly high $L_{\rm bol}/L_{\rm Edd}$ estimates, while the intermediate quasar SDSS J1447 does (Table \ref{tab:obslog}). However, we reiterate that these estimates may be highly uncertain, as noted in Section \ref{subsec:discuss_J0908}. \citet{Marlar18} included SDSS J1447 in a study investigating the X-ray spectra of a sample of WLQs; they found that it had a soft X-ray excess with a steep 0.5$-$8~keV power-law photon index ($\Gamma = 2.21^{+0.16}_{-0.15}$) and suggested this result was connected to high $L_{\rm bol}/L_{\rm Edd}$ (based on, e.g., \citealt{Done12}). SDSS J0908 was examined in X-ray prior to this by \citet{Luo15} and, notably, was also found to have a soft X-ray excess with a steep photon index, albeit with a somewhat larger uncertainty ($\Gamma = 2.2^{+0.5}_{-0.4}$). It is plausible that $L_{\rm bol}/L_{\rm Edd}$ is simply underestimated for SDSS J0908.

For completeness, we briefly evaluate our targets in the contexts of alternative scenarios for soft ionizing SEDs. Our spectral-index measurements from HST seem to disfavor the exceptionally massive SMBH scenario \citep{Laor11}. We measure $\alpha_{\rm NUV}$ and $\alpha_{\rm FUV}$ typical of normal radio-quiet quasars (see Section \ref{subsec:results_absmag}), while \citet{Laor11} predict steeply-falling spectra at $\lambda < 1000$~{\AA}. There is also the possibility that intrinsic X-ray weakness from a cold accretion disk is responsible for a soft ionizing SED, as with the weak-lined PHL 1811 \citep[][]{Leighly07a,Leighly07b}. This interpretation is unlikely to apply to WLQs, however, because: (1) on average, X-ray weak WLQs show signs of X-ray absorption or obscuration \citep[e.g.,][]{Wu12,Luo15}; (2) $\sim$50\% of the WLQ population is still X-ray normal \citep[e.g.,][]{Luo15,Ni18,Pu20}; and (3) there does not appear to be a correlation between X-ray weakness and emission-line weakness in the WLQ/intermediate populations (Section \ref{subsec:results_x-ray}). Indeed, our most securely-confirmed WLQ, SDSS J0908, is X-ray normal ($\Delta\alpha_{\rm ox}=0.11$) and is therefore contrary to the intrinsic X-ray weakness scenario.

\subsection{Intrinsic NALs and BALs in Weak-Lined Quasars} \label{subsec:discuss_absorp}

There is mounting evidence suggesting that outflow components such as disk winds are generally responsible for intrinsic NAL and BAL features in quasars \citep[e.g.,][]{Culliton19,Yi19b,Lu19,Rankine20}, and that some observed properties of wind-dominated quasars, such as \ion{C}{4} emission line blueshift, may arise from the same outflow systems \citep{Yi20}. Recently, \citet{Yi19a} observed a \ion{Mg}{2} BAL (i.e., ``LoBAL," displaying both low- and high-ionization absorption lines) WLQ with prominent BALs that disappeared and emerged over a period of $\sim$6 years. They suggested this activity was the result of transverse motion of outflowing gas, likely associated with a disk wind. There may be an overlap between wind-dominated WLQs and intrinsic NAL/BAL quasars that has yet to be explored in detail, as most WLQ studies exclude objects with clear signs of BALs to ensure that weak emission features are intrinsic and X-ray results are more straightforwardly interpretable.

Of our targets, SDSS J0812 and SDSS J1447 display signatures of intervening NAL systems, tentatively identified in the SDSS quasar \ion{Mg}{2} absorber catalog of \citet{Seyffert13}, and now confirmed via identification of additional, shorter-$\lambda$ absorption lines in the HST spectra (see Table \ref{tab:windows}). These systems are offset blueward from the objects' systemic redshifts by $v_{\rm off} > 100,000$ km s$^{-1}$, such that it is unlikely they are either intrinsic (i.e., physically related to outflows from the central engine) or associated (i.e., observed at the quasar systemic redshift) \citep[e.g.,][]{Misawa07,Shen12a}.

We have also tentatively identified in SDSS J1252 a narrow absorption system, with a blueward offset $v_{\rm off} \approx 14,400$ km s$^{-1}$, that has not been previously reported. Notably, this system lacks an identifiable \ion{Mg}{2} $\lambda2800$ absorption doublet in the SDSS spectrum. The remaining two objects, SDSS J0908 and SDSS J1530, appear to contain narrow absorption features in their HST spectra, and while we cannot identify specific systems, they also do not show corresponding narrow absorption in their SDSS spectra. While this result may be $-$ at least in part $-$ an effect of the SDSS spectral resolution, we cannot discount the possibility that these absorption systems are variable. Intervening NALs are not seen to vary greatly over decade scales, but variability and moderately high velocity offsets are two of the hallmarks expected of \textit{intrinsic} NALs originating from quasar outflows such as disk winds \citep[e.g.,][and references therein]{Culliton19}.

While more detailed characterization of these systems is beyond the scope of this paper, an additional feature in the spectrum of SDSS J1252 bears further comment. From the identified NAL system, we expect to see a \ion{Mg}{2} $\lambda1240$ absorption doublet centered on $\lambda$\textsubscript{rest} $\approx 1212.5$~{\AA}. Indeed, a strong absorption feature is identified here (see Figure \ref{fig:all_lyalpha}). However, this \ion{Mg}{2} transition is expected to be relatively weak based on its oscillator strength ($f \lesssim 0.0006$; \citealt{Kelleher08,Danforth16}, and references therein), and we suspect that the observed feature is instead Ly$\alpha$ from a different, associated absorption system (at $v_{\rm off} \approx 750$ km s$^{-1}$ as measured from 1216~{\AA} to the center of the trough). Based on visual examination of the observed trough width, we estimate ${\rm FWHM}\approx1970$ km s$^{-1}$ and tentatively propose that SDSS J1252 is a mini-BAL quasar.\footnote{Note that the absorption profile appears borderline saturated and/or damped, such that we may instead be seeing a proximate sub-damped Ly$\alpha$ system \citep[e.g.,][and references therein]{Noterdaeme19}. However, it is also possible that the profile simply consists of multiple narrow absorption lines that cannot be resolved with the available spectral resolution \citep[e.g.,][]{Lu19}.}

The above results and interpretations imply there may be physical qualities of BAL, mini-BAL, and/or intrinsic NAL quasars that relate to WLQs, but they also hint at weaker, possibly variable, intrinsic absorption systems being a more significant UV-band contaminant in $z<3$ WLQs or intermediate quasars than indicated by SDSS-based surveys of the general quasar population. For example, LoBAL quasars have a tendency to possess redder continua, stronger \ion{Fe}{2} emission, and comparatively weak \ion{C}{3}] emission in tandem with typical X-ray weakness; this behavior is commonly attributed to a larger viewing angle into the central engine \citep[e.g.,][]{Weymann91,Voit93}. Qualitatively, some X-ray weak WLQs and intermediate quasars (such as SDSS J1252) also display these features in contrast to their X-ray normal counterparts \citep{Wu12,Luo15}. If WLQs arise through the slim-disk scenario (in which X-ray weakness and our likelihood of looking ``through" winds are both expected to increase with inclination angle), and if WLQ winds are physically similar to those driving BALs and/or intrinsic NALs (e.g., if they are both linked to higher $L_{\rm bol}/L_{\rm Edd}$), then intrinsic absorption at moderate velocity offsets might be more observationally common in X-ray weak WLQs than X-ray normal WLQs (see \citealt{Rivera20} and \citealt{Richards21} for similar discussions in regard to associated absorption lines at the systemic redshift).

Several aspects of WLQ central engines need further exploration before confidently connecting WLQ winds with those driving BALs or intrinsic NALs. For example, the extent to which X-ray variability is possible across the full WLQ population needs characterization \citep{Ni20}, as does the likelihood of the BAL condition itself contributing to observed X-ray weakness \citep[e.g.,][]{Green95,Gallagher02}. This question would benefit from multi-epoch X-ray and optical$-$UV observations (in particular, higher-resolution UV spectra) of verified low-$z$ WLQs, with an emphasis on searching for variable, associated BALs/NALs in a wind-dominated, X-ray weak sample \citep[e.g.,][]{Yi19a}.

\section{Summary and Conclusions} \label{sec:summ}

We have presented HST STIS UV spectroscopy covering the Ly$\alpha$ + \ion{N}{5} complex for six candidate WLQs at low redshift ($0.9 < z < 1.5$). Combined with the objects' SDSS spectra, these observations allow direct comparison of high- and low-ionization emission species in individual objects. The HST targets possess UV continuum slopes and luminosities typical of normal radio-quiet quasars. While all six quasars display relatively weak \ion{Mg}{2} emission in their SDSS spectra, the new constraints on their Ly$\alpha$ + \ion{N}{5} emission qualify only two (SDSS J0812 and SDSS J0908) as \textit{bona fide} WLQs under the definition introduced by \citetalias{DS09}. Additionally, we obtain a constraint on weak \ion{C}{4} emission for SDSS J0908 that further secures its classification as a WLQ. One target (SDSS J1629) appears heavily reddened in its HST spectrum but otherwise shows Ly$\alpha$ + \ion{N}{5} emission similar to normal radio-quiet quasars, so it is excluded from analysis. The remaining three targets (SDSS J1252, SDSS J1447, and SDSS J1530) still have somewhat weak Ly$\alpha$ + \ion{N}{5} emission and likely represent the intermediate part of the population between WLQs and typical radio-quiet quasars (see \citealt{Ni18}). All HST targets were initially selected on the basis of weak \ion{Mg}{2} emission, but the observed range of Ly$\alpha$ EW further supports previous conclusions that weak low-ionization lines do not guarantee exceptionally weak high-ionization lines (e.g., \citealt{Shemmer10}; \citealt{Wu12}; \citetalias{Plotkin15}; \citealt{Luo15}).

The WLQ SDSS J0908 is more than $3\sigma$ weaker in \ion{C}{4} than expected from the Modified Baldwin Effect (e.g., \citealt{Shemmer15}), while \citetalias{Plotkin15} found that SDSS J1447 is only $\sim$1.5$\sigma$ weaker than expected and predicted (correctly) that it does not meet the weak EW[Ly$\alpha$+\ion{N}{5}] threshold. We tentatively (with caveats discussed in Section \ref{subsec:discuss_J0908}) suggest that, in the absence of Ly$\alpha$ coverage, it may be useful to verify that a quasar's EW[\ion{C}{4}] is significantly weaker than predicted by the Modified Baldwin Effect before classifying it as a WLQ.  This proposal requires additional validation, however, which may be possible by obtaining a sample of low-$z$ WLQ candidates with coverage of both Ly$\alpha$ and \ion{C}{4} as well as a reliable constraint on $L_{\rm bol}/L_{\rm Edd}$ from multiple estimate techniques (e.g., virial black-hole mass, $\Gamma-L_{\rm bol}/L_{\rm Edd}$ relationship, and/or \ion{C}{4} distance$-L_{\rm bol}/L_{\rm Edd}$ relationship).

We have compared the strengths of high- vs.\ low-ionization emission lines in our targets via the ratio $R_{\rm Ly\alpha,MgII} =$ EW[Ly$\alpha$+\ion{N}{5}]$/$EW[\ion{Mg}{2}] and evaluated them in the contexts of various proposed models for BELR weakness. Our targets display a range of comparatively small $R_{\rm Ly\alpha,MgII}$ with preferentially weak Ly$\alpha$, which favors a soft ionizing SED scenario. Furthermore, the UV and X-ray properties of our targets (including a lack of correlation found between EW[Ly$\alpha$+\ion{N}{5}] and X-ray weakness) suggest they are not intrinsically X-ray weak, and they appear to be most consistent with the ``slim-disk" shielding gas scenario \citep[e.g.,][]{Wu11,Wu12,Luo15,Ni18,Ni22}. However, we still cannot dismiss the possibility that WLQs represent a heterogeneous population with multiple mechanisms contributing toward BELR weakness (e.g., \citetalias{Plotkin15}). Additionally, an open issue remains in that the slim-disk model emphasizes high $L_{\rm bol}/L_{\rm Edd}$, while existing \ion{Mg}{2}-based estimates for our bona fide WLQs are not particularly high. The X-ray properties of SDSS J0908 suggest that $L_{\rm bol}/L_{\rm Edd}$ may be underestimated, but this remains to be verified.

Finally, we unexpectedly found evidence of NAL systems in several of the HST spectra, despite little to no sign of their presence in the corresponding SDSS spectra. While the majority of these appear to be intervening systems, we find tentative indications that a NAL system observed in SDSS J1252 may be intrinsic (i.e., physically related to outflows from the central engine). We have also observed a stronger absorption feature overlapping the Ly$\alpha$ emission of SDSS J1252, so we suspect this object may be a mini-BAL quasar. These findings hint at relatively weak absorption being more of a contaminant in the candidate WLQ population than previously thought, posing a challenge to low-$z$ WLQ identification. Given that recent observations of both BALs and associated NALs in normal quasars and WLQs have suggested a wind-driven source or component of absorption \citep[e.g.,][]{Culliton19,Lu19,Yi19a,Yi20}, we speculate that in the context of the orientation-dependent, slim-disk shielding gas scenario, intrinsic (though possibly variable) BALs and/or NALs may be more common in X-ray weak WLQs than X-ray normal WLQs. This result provides not only more avenues for comparison between WLQ and normal radio-quiet quasar populations, but also further potential insight into general quasar properties.

\acknowledgments{
We thank the anonymous referee for thoughtful and constructive feedback that helped improve this manuscript. J.D.P. thanks Roberto Mancini for advice on absorption measurement. W.N.B. thanks Chandra X-ray Center grant GO0-21080X and the V.M. Willaman Endowment at Penn State. B.L. acknowledges financial support from the National Natural Science Foundation of China grant 11991053, China Manned Space Project grants NO. CMS-CSST-2021-A05 and NO. CMS-CSST-2021-A06. Based on observations made with the NASA/ESA Hubble Space Telescope, obtained at the Space Telescope Science Institute, which is operated by the Association of Universities for Research in Astronomy, Inc., under NASA contract NAS5-26555. These observations are associated with Program number HST-GO-13298.001. Support for Program number HST-GO-13298.001 was provided by NASA through a grant from the Space Telescope Science Institute, which is operated by the Association of Universities for Research in Astronomy, Incorporated, under NASA contract NAS5-26555. Funding for the SDSS and SDSS-II has been provided by the Alfred P. Sloan Foundation, the Participating Institutions, the National Science Foundation, the U.S. Department of Energy, the National Aeronautics and Space Administration, the Japanese Monbukagakusho, the Max Planck Society, and the Higher Education Funding Council for England. The SDSS Web Site is http://www.sdss.org/. For this research, we made use of the SIMBAD database and VizieR catalogue access tool, operated at CDS, Strasbourg, France, as well as the NASA/IPAC Extragalactic Database (NED), which is funded by the National Aeronautics and Space Administration and operated by the California Institute of Technology. We also used the \texttt{Python} language along with \texttt{Astropy} \citep{astropy13, astropy18}, \texttt{NumPy} \citep{Numpy}, \texttt{SciPy} \citep{scipy}, and \texttt{TOPCAT} \citep{topcat}.
}

\appendix
\restartappendixnumbering

\section{Notes on Individual Objects} \label{sec:app_indiv}

\subsection{SDSS J0812} \label{subsec:J0812}

With EW[Ly$\alpha$+\ion{N}{5}] $= 15.3 \pm 2.3$~{\AA}, this object qualifies as a WLQ according to the \citetalias{DS09} criterion. While the 1$\sigma$ uncertainty could plausibly push it above the 15.4~{\AA} threshold, we reiterate that this boundary should not be considered fixed (Section \ref{sec:results}; see also Section 2 of \citetalias{DS09}). A spectrum covering \ion{C}{4} (which unfortunately falls within the gap between the SDSS and our HST spectra) would be useful to help confirm its WLQ classification.

While we confirm the presence of an intervening NAL system for this object at $z=0.79$ (originally identified in the \ion{Mg}{2} absorber catalog of \citealt{Seyffert13}), there is no indication this system influences the observed flux within the Ly$\alpha$ measurement range, nor is there any evidence of broad absorption troughs in either the HST or SDSS spectra. We conclude that this object is likely intrinsically weak-lined.

\subsection{SDSS J0908} \label{subsec:J0908}

We classify this object as a WLQ, with EW[Ly$\alpha$+\ion{N}{5}] $= 10.7 \pm 2.3$~{\AA} and EW[\ion{C}{4}] $= 8.3^{+3.8}_{-6.5}$~{\AA}. Its \ion{C}{4} emission is more than $3\sigma$ weaker than anticipated from the Modified Baldwin Effect (based on its $L_{\rm bol}/L_{\rm Edd}$; e.g., \citealt{Shemmer15}; with the caveat of virial black-hole mass estimates discussed in Section \ref{subsec:discuss_J0908}), and is also blueshifted ($\Delta v$[\ion{C}{4}] $= 1742 \pm 249.6$ km s$^{-1}$). While this WLQ is X-ray normal ($\Delta \alpha_{\rm ox} = 0.11$), its $\Delta \alpha_{\rm ox}-$EW[\ion{C}{4}] relationship is within the observed range for WLQs \citep[e.g.,][]{Ni18,Ni22,Timlin20}.

While there are possible NAL features in the HST spectrum at short wavelength, no system is identified, and the Ly$\alpha$ measurement range does not appear to be affected. Since there is also no evidence of broad absorption troughs, this object is likely intrinsically weak-lined.

\subsection{SDSS J1252} \label{subsec:J1252}

Despite not showing any significant absorption features in its SDSS spectrum, this quasar displays a NAL system at $z=1.24$ in its HST spectrum, as well as a strong, possibly associated Ly$\alpha$ absorber (see Section \ref{subsec:discuss_absorp}). Consequently, we suspect this may be a mini-BAL quasar, although further analysis in this regard is outside the scope of this work. Our EW measurement (EW[Ly$\alpha$+\ion{N}{5}] $= 21.9 \pm 2.7$~{\AA}) may constitute a lower limit, but it nevertheless clearly excludes SDSS J1252 from classification as a WLQ.

\subsection{SDSS J1447} \label{subsec:J1447}

SDSS J1447 was studied extensively in the rest-frame optical$-$UV range by \citetalias{Plotkin15} using spectra from the X-shooter spectrograph \citep{Vernet11} on the Very Large Telescope, producing line measurements for key emission features including H$\alpha$, H$\beta$, \ion{Mg}{2}, and \ion{C}{4}. Our HST observation extends rest-frame UV spectral coverage of this quasar blueward from $\sim$1450~{\AA} to $\sim$675~{\AA}, allowing comparison with Ly$\alpha$. Despite SDSS J1447's mildly weak \ion{C}{4} emission, the strength of its Ly$\alpha$ emission (EW[Ly$\alpha$+\ion{N}{5}] $= 26.4 \pm 3.0$~{\AA}) excludes it from classification as a WLQ. This confirms the conclusion reached by \citetalias{Plotkin15}, who noted that because its \ion{C}{4} emission was only $\sim$1.5$\sigma$ weaker than expected for its $L_{\rm bol}/L_{\rm Edd}$ and displayed relatively low blueshift, it was likely not a WLQ (instead suggesting it may be an intermediate quasar).

We have confirmed the presence of an intervening NAL system (originally identified by \citealt{Seyffert13}) at $z=1.03$ for this quasar. This system includes a Lyman edge visible near the blue limit of the HST spectrum. The broad Ly$\alpha$ emission line may experience some absorption (evidenced by what appears to be a prominent \ion{N}{5} $\lambda 1243$~{\AA} remnant; Figure \ref{fig:all_lyalpha}), although for the present we do not attempt to characterize this feature.

\subsection{SDSS J1530} \label{subsec:J1530}

This object is excluded from WLQ qualification based on the measurement of EW[Ly$\alpha$+\ion{N}{5}] $= 28.3 \pm 3.2$~{\AA}. It displays no detectable NAL features in its SDSS spectrum, and although there may be possible NALs in its HST spectrum, we are unable to identify a system.

\subsection{SDSS J1629} \label{subsec:J1629}

This object displays a combined (HST+SDSS) spectrum that is redder than typical, and we have measured a positive-valued FUV continuum slope (i.e., $\alpha_{\rm FUV} > 0$). It is not possible to find a broken power law fit that aptly represents the entire observed NUV range, so we measure the NUV continuum slope ($\alpha_{\rm NUV}$) for this quasar using the SDSS spectrum. Its measurement of EW[Ly$\alpha$+\ion{N}{5}] $= 83.7 \pm 18.6$~{\AA} is essentially normal (compared to, e.g., the ``typical" $92.91 \pm 0.72$~{\AA} from the composite of \citealt{VB01}), clearly excluding it from classification as a WLQ, and its $L_{\rm bol}/L_{\rm Edd}$ estimate (with the caveat of virial black-hole mass estimates; see Section \ref{subsec:discuss_J0908}) is an order of magnitude smaller than in our other targets. Based on these characteristics, we have largely excluded this quasar from Sections \ref{sec:results} and \ref{sec:discussion}.

\citet{Luo15} demonstrated that this quasar is neither X-ray weak ($\Delta \alpha_{\rm ox} = -0.02$) nor particularly red in its SDSS spectrum (relative color $\Delta(g-i)=0.01$), yet it appears to have a significantly harder X-ray hardness ratio than most of the WLQ candidates they observed (see their Table 1). Given the appearance of its HST spectrum and the lack of a soft X-ray excess, we suspect it is simply highly absorbed across the UV to soft X-ray band. It is also intriguing that the observed HST flux does not agree well with the GALEX NUV apparent magnitude (see \citealt{Luo15}, panel 3 of the extended online version of their Fig. 11, for a visual of the SED incorporating GALEX $m_{\rm NUV}$), further indicating that absorption (or perhaps variability) may be involved. In the optical$-$UV range, it appears the onset of the absorption unfortunately falls within the gap between the HST and SDSS spectra. Combined with this quasar's somewhat weak \ion{Mg}{2} emission, the above results leave the exact state of its ionizing continuum and its physical differences with WLQs and intermediate quasars in question. These questions may yet be useful as drives for future observations of larger samples.

\bibliography{WLQ2020.bib}{}
\bibliographystyle{aasjournal}

\end{document}